\documentclass[3p,times]{elsarticle}

\usepackage{amsmath,amssymb}

\usepackage{graphicx,color}

\usepackage{amsthm}

\usepackage[colorlinks=true]{hyperref}

\biboptions{sort&compress}

\usepackage{url}

\usepackage{bbm}

\renewcommand{\vec}[1]{\mathbf{#1}}

\newcommand{\ii}{\mathrm{i}}

\numberwithin{equation}{section}	

\theoremstyle{plain}			

\theoremstyle{definition}		

\DeclareMathOperator{\tr}{tr}

\journal{J. Phys. A}

\begin{document}

\begin{frontmatter}

\title{Efficient computation of the $W_3$ topological invariant and application to Floquet-Bloch systems}

\author{B. H{\"o}ckendorf}
\author[aa]{A. Alvermann\corref{cor1}}
\ead{alvermann@physik.uni-greifswald.de}
\author{H. Fehske}

\address{Institut f\"ur Physik, Ernst-Moritz-Arndt-Universit\"at Greifswald, 17487 Greifswald, Germany}

\begin{abstract}
 We introduce an efficient algorithm for the computation of the $W_3$ invariant of general unitary maps,
 which converges rapidly even on coarse discretization grids.
 The algorithm does not require extensive manipulation of the unitary maps,
 identification of the precise positions of degeneracy points, or fixing the gauge of eigenvectors.
After construction of the general algorithm,
we explain its application  to the $2+1$ dimensional maps that arise in the Floquet-Bloch theory of periodically driven two-dimensional quantum systems.
We demonstrate this application by
computing the $W_3$ invariant for an irradiated graphene model with a continuously modulated Hamilton operator, where it predicts the number of anomalous edge states in each gap. 
\end{abstract}


\end{frontmatter}

\section{Introduction}
\label{sec:Intro}

Topological invariants have gained considerable interest in solid state physics and related fields
through their application to the quantum Hall effect~\cite{PhysRevLett.45.494,PhysRevLett.49.405}, topological insulators~\cite{PhysRevLett.95.146802,RevModPhys.82.3045,Konig766,PhysRevLett.98.106803}, and more recently, Floquet-Bloch systems~\cite{Floquettopological,PhysRevB.84.235108,Flaschner,Wang453,PhysRevB.82.235114,PhysRevX.3.031005,Zhou2014,PhysRevLett.112.026805,%
PhysRevB.90.195419,%
PhysRevLett.109.010601,%
PhysRevB.93.075405,%
PhysRevLett.114.106806,%
1367-2630-17-12-125014%
}.
While the relevant topological invariants for the quantum Hall effect are the Chern numbers of the respective bands of the Hamiltonian, the invariants for Floquet-Bloch systems are constructed for unitary maps that are derived from the time propagator of the periodically driven system. These invariants, which can be defined generally for unitary maps in odd dimensions~\cite{Naka,Wein,
Bott1978,%
Prodan20161150,%
PSB16,%
2017arXiv170107455L%
}, can be understood as the generalization of the winding number of a circle map $\mathbb S^1 \to \mathbb S^1$.
For Floquet-Bloch systems in two spatial and one time dimension, the $W_3$ invariant is of primary interest. In particular, it gives the number of (anomalous) edge states that can not be predicted with other invariants such as the simpler $W_1$ invariant, which is directly related to the winding number of the eigenvalues of a unitary map,
or the Chern numbers of the Floquet bands.

In this paper we develop an algorithm for the efficient computation of the $W_3$ invariant of general unitary maps, and then provide the application to Floquet-Bloch systems.
Our algorithm is motivated by the algorithm of Fukui, Hatsugai, Suzuki~\cite{FHS05},
which allows for the direct gauge-invariant computation of Chern numbers.
However, while the Chern numbers depend only on the eigenvectors of the respective bands,
we must consider both the variation of eigenvalues and eigenvectors of the unitary map.
Furthermore, we cannot assume a gap condition: A non-zero $W_3$ invariant
implies the existence of degeneracy points of the unitary map.
Therefore, construction of our algorithm requires a few additional considerations.

The paper is organized as follows.
In Sec.~\ref{sec:W3} we specify the mathematical setting for this paper and fix our notation.
We start our construction of the algorithm in Sec.~\ref{sec:NoGauge} with the derivation of a gauge-invariant expression for the $W_3$ invariant that involves the eigenvalue and eigenvector bands of the unitary map, before we explain the algorithmic scheme in Sec.~\ref{sec:Algo}.
In Sec.~\ref{sec:Floquet} we address the application to Floquet-Bloch systems introduced in Ref.~\cite{PhysRevX.3.031005}, which involves only a small modification of the general computational scheme. The actual application to the anomalous edge states in an irradiated graphene model is given in Sec.~\ref{sec:Graphene}.
We conclude in Sec.~\ref{sec:Conc}.

\section{The $W_3$ invariant}
\label{sec:W3}

The mathematical setting for this paper is as follows.
Let $U(\cdot): \mathbb T^3 \to \mathbb U(n)$ denote a smooth map
from the three-dimensional torus $\mathbb T^3 = \mathbb S^1 \times \mathbb S^1 \times \mathbb S^1$ to the Lie group $\mathbb U(n)$ of unitary $n\times n$ matrices.
We can view $U(\cdot)$ as a map defined on the cube $[0,1]^3$ that is periodic in every argument, as in
\begin{equation}\label{BoundaryTorus}
U(0,\mu_2, \mu_3) = U(1, \mu_2,\mu_3) \;,
\quad
U(\mu_1, 0, \mu_3) = U(\mu_1, 1, \mu_3) \;,
\quad
U(\mu_1, \mu_2, 0) = U(\mu_1,\mu_2, 1) \;.
\end{equation}
In this situation, the quantity $W_3[U]$ is defined as~\cite{PhysRevB.82.235114,PhysRevX.3.031005}
\begin{equation}\label{W3}
 W_3[U] = \frac{1}{24 \pi^2 } \int\limits_{[0,1)^3} \epsilon_{\alpha\beta\gamma} \tr \big[ (U^{-1} \partial^\alpha U) \, (U^{-1} \partial^\beta U) \, (U^{-1} \partial^\gamma U)   \big]   \, \mathrm d^3 \mu\;.
\end{equation}
Here, $\epsilon_{\alpha\beta\gamma}$ denotes the antisymmetric Levi-Civita symbol.
We write $\partial^\alpha A(\mu) = \partial A(\mu)/\partial \mu_\alpha$, with $\alpha = 1,2,3$, for the derivative of a matrix-valued function $A(\cdot)$.
The indices $\alpha, \beta, \gamma$ are counted modulo 3. For example,
if one coordinate $\mu_\alpha$ is given, the remaining two coordinates are
$\mu_{\alpha-1}$ and $\mu_{\alpha+1}$.
Summation over repeated indices $\alpha, \beta, \gamma$ is implied,
and we often drop the argument $\mu$ to lighten the notation.

$W_3[U]$ is an integer that is invariant under continuous deformations of $U(\cdot)$, i.e., it is a topological invariant.
Note that $W_3[U]=0$ if $U(\cdot)$ is constant along one direction $\mu_\alpha$.
In addition to $W_3[U]$ we have three invariants
\begin{equation}
 W^\alpha_1[U] = \frac{1}{2 \pi \ii} \int\limits_0^1 \tr [ U^{-1} \partial^\alpha U ] \, \mathrm d \mu_\alpha 
\end{equation}
along each direction $\alpha=1,2,3$.
These invariants are directly related to the winding numbers of the eigenvalues of $U(\cdot)$,
while $W_3[U]$ also involves the eigenvectors as we will see below.
Note that, for a smooth map, $W^\alpha_1[U]$ does not depend on the values of the two coordinates $\mu_{\alpha-1}$, $\mu_{\alpha+1}$ kept fixed in the above expression.

\section{Gauge-invariant expression for the $W_3$ invariant}
\label{sec:NoGauge}

Let us assume that $U(\cdot)$ is given by its decomposition into eigenvalues and eigenvectors,
\begin{equation}\label{USD}
U(\mu) = S(\mu) D(\mu) S^\dagger(\mu) \;,
\end{equation}
with a unitary matrix map $S(\cdot)$ and a diagonal matrix map $D(\cdot)$.
The columns of $S(\mu)$ give the eigenvectors $\vec s^\nu(\mu)$ of $U(\mu)$,
while the corresponding eigenvalues $d^\nu(\mu)$ appear as the entries of $D(\mu)$.

Since the above decomposition is not unique, the numbering of the eigenvalues and eigenvectors can affect the analytical properties of the maps $S(\cdot)$, $D(\cdot)$. 
In contrast to the Hermitian case, there is also no natural ordering of the eigenvalues $d^\nu(\mu) \in \mathbb S^1$.
In our algorithm, the indices $\nu$ will be assigned according to the distance of eigenvalues and the overlap of eigenvectors on adjacent discretization points (see below).

Away from degeneracy points of $U(\cdot)$, perturbation theory~\cite{Kato66} implies that $S(\cdot)$, $D(\cdot)$ can be constructed as smooth maps, essentially by assigning a fixed index $\nu$ to each of the separated `bands' of $U(\cdot)$.
In the vicinity of a degeneracy point of $U(\cdot)$, where two eigenvalues coincide,
the situation depends on the dimensionality of the domain of $U(\cdot)$.
In one dimension, for $\mu \in \mathbb R$, it would be possible to continue the maps $S(\cdot)$, $D(\cdot)$ smoothly through the degeneracy point.
In dimensions two or higher, as considered here, only continuity of the eigenvalues (but not smoothness) can be achieved in the general case.
The eigenvectors may fail to be even continuous at the degeneracy point, as witnessed by the second example in Sec.~\ref{sec:Example}.
For a generic map $U(\cdot)$ that is derived from a (possibly time-dependent) Hamiltonian, degeneracies occur at isolated points~\cite{Wigner}.
Therefore, we will assume that the maps $S(\cdot)$, $D(\cdot)$ are smooth,
apart from the isolated degeneracy points of $U(\cdot)$.

From the decomposition~\eqref{USD} we get
\begin{equation}\label{DUXYZ}
U^{-1} \partial^\alpha U = S \big( Z^\alpha +Y^\alpha - X^\alpha  \big) S^\dagger \;,
\end{equation}
with
\begin{equation}
 X^\alpha = S^\dagger \partial^\alpha S \;, \quad Y^\alpha = D^\dagger \partial^\alpha  D = (\partial^\alpha  D) D^\dagger \;,
 \quad Z^\alpha = D^\dagger X^\alpha D \;.
\end{equation}
Note that $D^\dagger$ and $\partial^\alpha D$ commute,
and that $(X^\alpha)^\dagger = - X^\alpha$, $(Y^\alpha)^\dagger = - Y^\alpha$,
$(Z^\alpha)^\dagger = - Z^\alpha$.
With these definitions, we get
\begin{equation}
\tr \big[ (U^{-1} \partial^\alpha  U) \,  (U^{-1} \partial^\beta U) \, (U^{-1} \partial^\gamma U) \big]
= \tr \big[ ( Z^\alpha +Y^\alpha - X^\alpha  )  (Z^\beta +Y^\beta - X^\beta   )  ( Z^\gamma  +Y^\gamma - X^\gamma  ) \big] \;,
\end{equation}
and so
\begin{equation}\label{TrXXX}
\begin{split}
& \epsilon_{\alpha\beta\gamma} \tr \big[ (U^{-1} \partial^\alpha  U) \,  (U^{-1} \partial^\beta U) \, (U^{-1} \partial^\gamma U) \big] = \\  
 &  \epsilon_{\alpha\beta\gamma} 
 \tr  \big[ 
 - X^\alpha X^\beta X^\gamma 
 + 3 X^\alpha X^\beta Y^\gamma 
+ 3 X^\alpha X^\beta Z^\gamma 
-3 X^\alpha Y^\beta Z^\gamma 
-3 X^\alpha Z^\beta Y^\gamma \\
& \qquad\qquad\qquad\qquad\qquad\qquad\qquad -3 X^\alpha Z^\beta Z^\gamma 
+3 Y^\alpha Z^\beta Z^\gamma 
 + Z^\alpha Z^\beta Z^\gamma   \big]
\end{split}
\end{equation}
for the integrand of Eq.~\eqref{W3}.
Here, we arrange products of operators in lexicographic order, using the cyclicity of $\epsilon_{\alpha\beta\gamma}$ and of the trace $\tr[\cdot]$.
Note that terms with two or more $Y$ factors cancel, because these matrices commute among themselves.
Note further that the $D$, $D^\dagger$ matrices in terms with $Z$ factors partially cancel such that, e.g., $\tr[Y^\alpha Z^\beta Z^\gamma] = \tr[Y^\alpha X^\beta X^\gamma]$.
In particular, the first ($X^\alpha X^\beta X^\gamma$) and last ($Z^\alpha Z^\beta Z^\gamma$) term in Eq.~\eqref{TrXXX} cancel.

From the derivatives
\begin{equation}
 \partial^\alpha S^\dagger = - S^\dagger (\partial^\alpha S) S^\dagger \;,
 \qquad
 \partial^\alpha D^\dagger = - D^\dagger (\partial^\alpha D) D^\dagger
\end{equation}
of the unitary  matrices $S(\mu)$, $D(\mu)$ we get the derivatives
\begin{subequations}\label{DerivativesXYZ}
\begin{align}
\partial^\alpha X^\beta &= - X^\alpha X^\beta + X^{\alpha \beta} \;, \\
\partial^\alpha Y^\beta &= - Y^\alpha Y^\beta + Y^{\alpha \beta} \;, \\
\partial^\alpha Z^\beta &= - Y^\alpha Z^\beta  - Z^\alpha Z^\beta + Z^{\alpha\beta} + Z^\beta Y^\alpha 
\end{align}
\end{subequations}
of the individual factors in the above equations, 
with $X^{\alpha\beta} = S^\dagger \partial^{\alpha\beta} S$,
$Y^{\alpha\beta} = D^\dagger \partial^{\alpha\beta} D$,
$Z^{\alpha\beta} = D^\dagger X^{\alpha\beta} D$,
and then the relations
\begin{subequations}
\begin{align}
 \partial^\alpha \big( Z^\beta X^\gamma \big) &= - Y^\alpha Z^\beta X^\gamma - Z^\alpha Z^\beta X^\gamma + Z^{\alpha \beta} X^\gamma + Z^\beta Y^\alpha X^\gamma - Z^\beta X^\alpha X^\gamma + Z^\beta X^{\alpha \gamma} \;, \\
 \partial^\alpha \big( Z^\beta Y^\gamma \big) &= - Y^\alpha Z^\beta Y^\gamma - Z^\alpha Z^\beta Y^\gamma + Z^{\alpha \beta} Y^\gamma + Z^\beta Y^{\alpha \gamma} \;.
\end{align}
\end{subequations}
These relations allow us to express Eq.~\eqref{TrXXX} in the form
\begin{equation}\label{Shabby1}
\epsilon_{\alpha\beta\gamma} \tr \big[ (U^{-1} \partial^\alpha  U) \,  (U^{-1} \partial^\beta U) \, (U^{-1} \partial^\gamma U) \big]
= \epsilon_{\alpha\beta\gamma} \Big( 3 \tr \big[ \partial^\alpha \big( Z^\beta X^\gamma \big)   \big]
 - 6  \tr \big[ \partial^\alpha \big( Z^\beta Y^\gamma \big) \big] \Big) \;.
\end{equation}
This expression suggests that one integration in Eq.~\eqref{W3}, over $\mathrm d\mu_\alpha$,
can be replaced by the difference of the boundary terms for $\mu_\alpha = 0, 1$,
and thus may cancel.

However, we cannot expect that the map $S(\cdot)$ in Eq.~\eqref{USD} or the factors $X^\alpha$, $Z^\alpha$ in Eq.~\eqref{DUXYZ} are compatible with the boundary conditions of the torus (as in Eq.~\eqref{BoundaryTorus}).
The reason is that the decomposition~\eqref{USD} is not unique but depends on the gauge of eigenvectors.
Under a gauge transformation $S(\cdot) \mapsto \tilde S(\cdot)$, which is defined by a map of unitary diagonal matrices $G(\cdot)$ as
\begin{equation}
 \tilde S(\mu) = S(\mu) G(\mu) \;,
\end{equation}
the above factors change as
\begin{equation}
 \tilde X^\alpha = G^\dagger X^\alpha G + G^\alpha \;,
 \quad
  \tilde Z^\alpha = G^\dagger Z^\alpha G + G^\alpha \;,
\end{equation}
with $G^\alpha = G^\dagger \partial^\alpha G$.
The quantity $Y^\alpha$ is gauge-invariant.

We must now compare the boundary values of the two terms in Eq.~\eqref{Shabby1}, which are related through a gauge transformation.
From $U(0,\mu_2,\mu_3) = U(1,\mu_2,\mu_3)$ we conclude that
$S(1,\mu_2,\mu_3) = S(0,\mu_2,\mu_3) G(\mu_2,\mu_3) P_1$
and $D(1,\mu_2,\mu_3) = P_1^\dagger D(0,\mu_2,\mu_3) P_1$,
with a diagonal gauge matrix $G(\mu_2, \mu_3)$ as before, and a permutation matrix $P_1 \equiv P_1(\mu_2,\mu_3)$. Because $P$ is a discrete quantity, it does not depend on $\mu_2$, $\mu_3$.
The permutation matrix cancels in all expression involving a trace, but the gauge $G$ does not.

Through the gauge transformation, the boundary values of the first term in Eq.~\eqref{Shabby1} are related by
\begin{equation}
 \tr\big[ Z^\beta X^\gamma \big] \big|_{\textstyle(1, \mu_2, \mu_3)} =  \tr\big[  Z^\beta X^\gamma +  
 Z^\beta G^\gamma + G^\beta X^\gamma + G^\beta G^\gamma 
 \big]\big|_{\textstyle(0, \mu_2, \mu_3)} \;,
\end{equation}
and similarly for the other directions. 
The last three terms cancel when combined with $\epsilon_{\alpha\beta\gamma}$
(note that $\tr\big[Z^\beta G^\gamma \big] = \tr\big[X^\beta G^\gamma \big]$),
and we see that the first term in Eq.~\eqref{Shabby1} does not contribute.

In the second term $\tr \big[ \partial^\alpha \big( Z^\beta Y^\gamma \big) \big] = \tr \big[ \partial^\alpha \big( X^\beta Y^\gamma \big) \big]$,
the factor  $X^\beta$ depends on the gauge of $S(\cdot)$ but $Y^\alpha$ does not.
The boundary values do not cancel and we have to keep this term.
Now, we perform the derivative and recognize that in $ \partial^\alpha \big( X^\beta Y^\gamma \big) = (\partial^\alpha X^\beta) Y^\gamma + X^\beta (-Y^\alpha Y^\beta + Y^{\alpha\beta})$ 
the second term cancels in the final expression because the $Y$ factors commute.
In this way, we obtain the manifestly gauge-invariant expression
\begin{equation}\label{W3FY}
W_3[U] = \frac{1}{2 \pi \ii} \int\limits_{[0,1)^3} \tr \big[ F_\alpha Y^\alpha \big] \, \mathrm d^3 \mu \;,
\end{equation}
with the Berry curvature matrix
\begin{equation}
 F_\alpha = \frac{1}{2\pi\ii} \epsilon_{\alpha\beta\gamma} (\partial^\beta X^\gamma) \;.
\end{equation}
This matrix is diagonal, and gauge-invariant by construction.

For future reference we note that $\partial^\alpha F_\alpha = 0$, unless two eigenvalues coincide at a degeneracy point of $U(\cdot)$. Furthermore, we always have $\tr\big[ F_\alpha \big] = 0$.

\subsection{Winding and Chern numbers}

We can view the decomposition~\eqref{USD} as a way to combine the eigenvalues and eigenvectors of $U(\cdot)$ into bands
$\mu \mapsto d^\nu(\mu)$ and $\mu \mapsto \vec s^\nu(\mu)$
for each $\nu=1, \dots, n$.
The corresponding entries of the diagonal matrices $Y^\alpha$, $F_\alpha$ in Eq.~\eqref{W3FY}
contain information about the winding and Chern numbers of these
bands.

The diagonal entries $Y^{\alpha,\nu}$ of $Y^\alpha$
give the angular velocity 
$- \ii Y^{\alpha,\nu}(\mu)$
of the $\nu$-th eigenvalue, as it moves on the circle $\mathbb S^1 \subset \mathbb C$.
Therefore, the winding number of band $\nu$ along direction $\alpha$ is
\begin{equation}\label{WindNu}
W^{\alpha, \nu} = \frac{1}{2 \pi \ii} \int\limits_0^1 Y^{\alpha,\nu}  \, \mathrm{d} \mu_\alpha \;,
\end{equation}
and the $W_1$ invariants of $U$ are given by
\begin{equation}
 W_1^\alpha [U] = \sum_{\nu=1}^n W^{\alpha, \nu} \;.
\end{equation}
$W^{\alpha, \nu}$ and $W^\alpha_1[U]$ are both integers.
Since the eigenvalues are smooth functions of $\mu$,
the value of $W^{\alpha, \nu}$ does not depend on the two coordinates $\mu_{\alpha-1}$, $\mu_{\alpha+1}$ kept fixed in Eq.~\eqref{WindNu}.

The diagonal entries $F^\nu_\alpha(\mu)$ of $F_\alpha$
give the Berry curvature of the $\nu$-th eigenvector.
Therefore, the integral
\begin{equation}\label{ChernNu}
C^\nu_\alpha = \int\limits_0^1 \!\!\! \int\limits_0^1 F^\nu_\alpha(\mu) \, \mathrm{d} \mu_{\alpha-1} \, \mathrm{d} \mu_{\alpha+1}
\end{equation}
gives the Chern number of band $\nu$ 
on the surface $\mu_\alpha = \mathrm{const.}$,
perpendicular to direction $\alpha$.
The value of $C^\nu_\alpha$ can change as a function of $\mu_\alpha$ when the integration surface passes through a degeneracy point of $U(\cdot)$, where the eigenvectors fail to be continuous.

\subsection{Global expression for the $W_3$ invariant}
\label{sec:Global}
Eq.~\eqref{W3FY} depends only on locally defined quantities. An alternative form of this equation can be obtained by introducing the globally defined eigenvalue bands of $U(\cdot)$. 
To this end, we note that the integrability condition $\epsilon_{\alpha\beta\gamma} \partial^\beta Y^\gamma = 0$ holds (see Eq.~\eqref{DerivativesXYZ}) 
such that the equation $\ii \partial^\alpha \Phi(\mu) = Y^\alpha(\mu)$
has a solution $\Phi(\cdot)$ defined on the entire cube $[0,1]^3$.
By construction, 
$\Phi(\mu)$ is a diagonal matrix with $D(\mu) = \exp[\ii \Phi(\mu)]$.
We can interpret the elements of this matrix as forming the bands $\mu \mapsto \Phi^\nu(\mu)$ of $U(\cdot)$.
Every band is defined only up to a global shift by an integer multiple of $2\pi$.
Especially at the boundaries $\mu_\alpha = 0,1$ of the cube,
where the eigenvalues of $U(\cdot)$ coincide,
we have $\Phi^\nu(\mu)\big|_{\mu_\alpha=1} =  \Phi^\nu(\mu)\big|_{\mu_\alpha=0} + 2 \pi W^{\alpha, \nu}$,
with the winding number of the $\nu$-th eigenvalue from Eq.~\eqref{WindNu}.
In particular, the difference between the $\Phi^\nu(\mu)$ is constant on each boundary face.

Partial integration of Eq.~\eqref{W3FY} gives
\begin{equation}\label{W3BigPhi}
W_3[U] =   \frac{1}{2 \pi} \left(  - \int\limits_{[0,1)^3} \tr \big[ (\partial^\alpha F_\alpha) \Phi \big]  \, \mathrm d^3 \mu \,
+ \, \sum_{\alpha=1}^3  \int\limits_0^1 \!\!\! \int\limits_0^1 \tr \big[F_\alpha \Phi \big] \Big|_{\mu_\alpha = 0}^{\mu_\alpha = 1} \, \mathrm{d} \mu_{\alpha-1} \, \mathrm{d} \mu_{\alpha+1}
\right) 
\;.
\end{equation}
The first term contains the divergence of $F^\alpha$ that can be non-zero only~\cite{Berry45} at degeneracy points of $U(\cdot)$.
The second term gives the boundary contributions at $\mu_\alpha = 0,1$.
This equation, which contains the global bands $\Phi(\cdot)$, is the counterpart to Eq.~\eqref{W3FY} that uses only local quantities derived from $U(\cdot)$.

\section{Computation of the $W_3$ invariant}
\label{sec:Algo}

\begin{table}
\hspace*{\fill}
\begin{tabular}{ccccc}
& $\vec p_1$ & $\vec p_2$ & $\vec p_3$ & $\vec p_4$ \rule[-9pt]{0pt}{2.6ex} \\\cline{2-5}
\rule{0pt}{2.6ex}
$\mathsf F_{\vec p,1}$: & $\vec p$ & $\vec p +  \boldsymbol\delta_2$  & $\vec p + \boldsymbol\delta_2 + \boldsymbol\delta_3$ & $\vec p + \boldsymbol\delta_3$ \\
\rule{0pt}{3ex}
$\mathsf F_{\vec p,2}$: & $\vec p$ & $\vec p + \boldsymbol\delta_3$  & $\vec p + \boldsymbol\delta_1 + \boldsymbol\delta_3$ & $\vec p + \boldsymbol\delta_1$ \\
\rule{0pt}{3ex}
$\mathsf F_{\vec p,3}$: & $\vec p$ & $\vec p + \boldsymbol\delta_1$  & $\vec p + \boldsymbol\delta_1 + \boldsymbol\delta_2$ & $\vec p + \boldsymbol\delta_2$
\end{tabular}
\hspace*{\fill}
\caption{Vertices $\vec p_1, \dots, \vec p_4$ of the faces $\mathsf F_{\vec p,\alpha}$ used in the algorithm.}
\label{tab:Face}
\end{table}

To compute $W_3[U]$ 
we use a grid of
$N \times N \times N$ discretization points 
$\vec p = (\delta i_1, \delta i_2, \delta i_3)$,
where $\delta = 1/N$ for some $N > 1$ and $1 \le i_\alpha \le N$.
To keep the presentation simple we assume an equidistant discretization grid,
but generalization of the following expressions to non-equidistant grids is straightforward.
In the following, $\boldsymbol\delta_\alpha$ denotes the lattice vector in direction $\alpha$,
i.e., $\boldsymbol\delta_1 = (\delta, 0 , 0)$ etc.,
and we count 
coordinates $\mu_\alpha$ modulo $1$, i.e., $\mu + N \boldsymbol\delta_\alpha = \mu$.
At every discretization point, we determine $S(\vec p)$ and $D(\vec p)$ through diagonalization of $U(\vec p)$.
We assume that, for given $\nu$, the eigenvalues $d^\nu(\vec p)$ and eigenvectors $\vec s^\nu(\vec p)$ on adjacent discretization points 
$\vec p$, $\vec p \pm \boldsymbol\delta_\alpha$
belong to the same band.
To achieve this in practice, we can compare the distance $|d^\nu(\vec p) - d^\nu(\vec p \pm \boldsymbol\delta_\alpha) |$ of the eigenvalues or match the eigenvectors according to their overlap $|\langle \vec s^\nu(\vec p), \vec s^\nu(\vec p \pm \boldsymbol\delta_\alpha) \rangle|$.
Note that we do not need to consistently assign the indices $\nu$ on the entire discretization grid, but only locally on each of the cubes.

\paragraph{Faces}
Every discretization point $\vec p$ is the base point of 
three rectangular faces $\mathsf F_{\vec p,\alpha}$,
with vertices as in Tab.~\ref{tab:Face}.
To every face $\mathsf F_{\vec p, \alpha}$ we assign the real numbers
\begin{equation}
 \hat F_{\vec p,\alpha}^\nu = \frac{1}{2\pi \ii} \log \mathcal U^\nu(\vec p_1, \vec p_2) \mathcal U^\nu(\vec p_2, \vec p_3) \mathcal U^\nu(\vec p_3, \vec p_4) \mathcal U^\nu(\vec p_4, \vec p_1)  \;,
\end{equation}
which are derived from the eigenvectors of $U(\cdot)$ via the gauge variables
\begin{equation}
\mathcal U^\nu(\vec p_i,\vec p_j) = \frac{\langle \vec s^\nu(\vec p_i) ,  \vec s^\nu(\vec p_j) \rangle}{ | \langle \vec s^\nu(\vec p_i) , \vec s^\nu(\vec p_j) \rangle|}
\end{equation}
defined along each edge of the face.
Here, $\langle \cdot , \cdot \rangle$ denotes the complex Euclidean scalar product,
and we can always avoid $\langle \vec s^\nu(\vec p_i) , \vec s^\nu(\vec p_j) \rangle = 0$ by making the discretization finer.
It is $ \mathcal U^\nu(\vec p_i,\vec p_j) \in \mathbb S^1$,
and $\hat F_{\vec p,\alpha}^\nu \in \mathbb R$.
For $\delta \to 0$, we have $\hat F_{\vec p, \alpha}^\nu =  F^\nu_\alpha (\vec p) \, \delta^2 + O(\delta^3)$.
This expression for $\hat F_{\vec p,\alpha}^\nu$ is identical to the construction in Ref.~\cite{FHS05}.

\begin{figure}
\hspace*{\fill}
\includegraphics[width=0.5\linewidth]{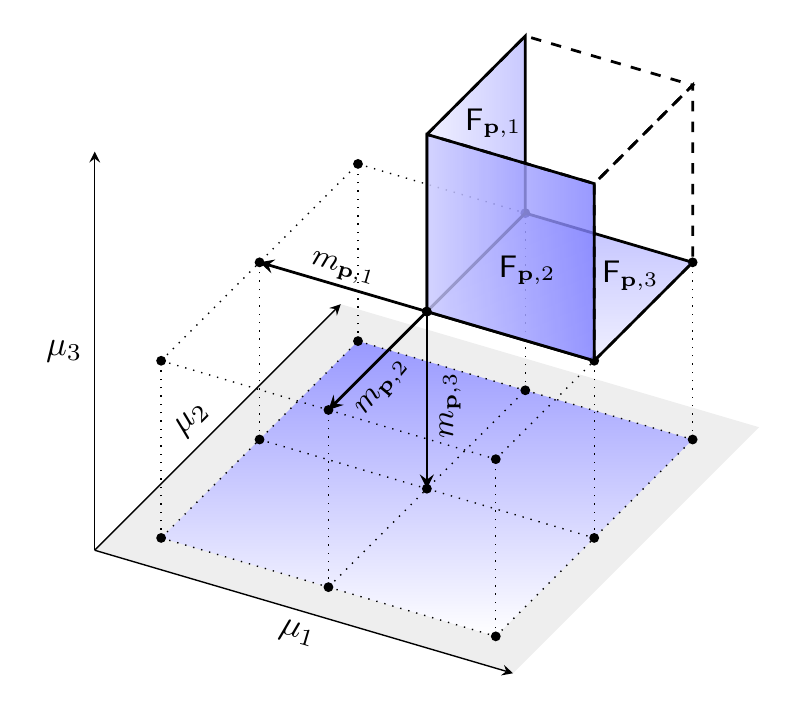}
\hspace*{\fill}
\caption{Sketch of the cube $\mathsf C_{\vec p}$ with the three faces $\mathsf F_{\vec p,\alpha}$,
and of the three edges $\vec p$ to $\vec p - \boldsymbol \delta_\alpha$ that determine the integers $m_{\vec p, \alpha}^\nu$ in the algorithm.
}
\label{fig:SketchFaces}
\end{figure}

\paragraph{Cubes}
Every discretization point $\vec p$ is also the base point of a cube $\mathsf C_{\vec p}$,
with opposite corner $\vec p + \boldsymbol\delta_1 + \boldsymbol\delta_2 + \boldsymbol\delta_3$ (see Fig.~\ref{fig:SketchFaces}).
To every cube we assign the number
\begin{equation}\label{Cpnu}
\hat C_{\vec p}^\nu = \sum_{\alpha=1}^3 \hat F_{\vec p + \boldsymbol\delta_\alpha, \alpha}^\nu
- \hat F_{\vec p, \alpha}^\nu \;.
\end{equation}
Every edge of the cube appears twice in this sum,
contributing a term $\mathcal U^\nu(\vec p_i,\vec p_j)$ and
$\mathcal U^\nu(\vec p_j,\vec p_i) = \mathcal U^\nu(\vec p_i,\vec p_j)^{-1}$.
Combining the logarithms in the sum thus results in $\exp\big(2 \pi \ii \hat C_{\vec p}^\nu \big)= 1$,
such that $\hat C_{\vec p}^\nu \in \mathbb Z$.
In the limit $\delta \to 0$,
$\hat C_{\vec p}^\nu$ equals the volume integral of $\partial^\alpha F_\alpha$ over the cube $\mathsf C_{\vec p}$,
and gives the Chern number of the $\nu$-th band of $U(\cdot)$ on an infinitesimal surface at $\vec p $.
Therefore, $\hat C_{\vec p}^\nu \ne 0$ is possible only if 
$\mathsf C_{\vec p}$ contains a degeneracy point $\mu_d$ of $U(\cdot)$,
where two eigenvalues 
$d^\nu(\mu_d) = d^{\nu'}(\mu_d)$
for different $\nu \ne \nu'$ coincide.
In this situation, we have 
$\partial^\alpha F^\nu_\alpha(\mu_d) = - \partial^\alpha F^{\nu'}_\alpha(\mu_d)$
in agreement with $\tr\big[ \partial^\alpha F_\alpha \big] = 0$,
and $\hat C_{\vec p}^\nu = - \hat C_{\vec p}^{\nu'}$ in agreement with $\sum_\nu \hat C_{\vec p}^\nu = 0$.

\paragraph{$\hat W_3$ approximation}
The approximation for the $W_3$ invariant computed by our algorithm is given by the sum
\begin{equation}\label{HatW3}
 \hat W_3 = \sum_{i_1, i_2, i_3 = 1}^N \sum_{\nu=1}^n  \, \left( \hat C_{\vec p}^\nu M_{\vec p}^\nu + \sum_{\alpha=1}^3   \hat F_{\vec p,\alpha}^\nu \, m_{\vec p, \alpha}^\nu \right)
 \end{equation}
over all faces and cubes,
where the integers $m_{\vec p, \alpha}^\nu$, $M_{\vec p}^\nu$ are obtained from the
logarithms
\begin{equation}\label{SmallPhi}
\phi_{\vec p}^\nu = - \ii \log d^\nu(\vec p)
\end{equation}
of the eigenvalues of $U(\cdot)$.
On the one hand, we determine $m_{\vec p, \alpha}^\nu$ such that $|\phi_{\vec p}^\nu - \phi_{\vec p - \boldsymbol \delta_\alpha}^\nu  + 2 \pi m_{\vec p, \alpha}^\nu| < \pi$.
Equality can be excluded by making the discretization finer.
On the other hand,
if $\hat C_{\vec p}^\nu = -  \hat C_{\vec p}^{\nu'} \ne 0$ for two
indices $\nu$, $\nu'$, we pick one index, for example $\nu$,
and determine $M_{\vec p}^\nu$ such that $| \phi_{\vec p}^\nu - \phi_{\vec p}^{\nu'}  + 2 \pi M_{\vec p}^\nu| < \pi$.
Only the term $\hat C_{\vec p}^\nu M_{\vec p}^\nu$  is included in the sum,
and we set $M_{\vec p}^{\nu'} = 0$.
If we pick the index $\nu'$ instead,
we have $M_{\vec p}^{\nu'} = - M_{\vec p}^\nu $ such that 
$ \hat C_{\vec p}^{\nu'} M_{\vec p}^{\nu'} =  \hat C_{\vec p}^\nu M_{\vec p}^\nu $.
Therefore, it does not matter which of the two terms is included in the sum~\eqref{HatW3}.

Note that the expression~\eqref{HatW3} is manifestly gauge-invariant, due to the construction of the $\hat F_{\vec p,\alpha}^\nu$.
Usually, only few faces and cubes with non-zero $m_{\vec p, \alpha}^\nu$, $M_{\vec p}^\nu$ contribute in the sum.

\paragraph{Winding and Chern numbers}

In addition to $\hat W_3$, the algorithm provides us with the approximation
\begin{equation}\label{HatW1}
 \hat W_1^\alpha = \sum_{\nu=1}^n \sum_{i_\alpha=1}^N m^\nu_{\vec p, \alpha}
\end{equation}
for the $W_1$ invariants of $U(\cdot)$.
By construction, $\hat W^\alpha_1$ is an integer,
and $\hat W_1^\alpha \to W^\alpha_1[U]$ for $\delta \to 0$.

Furthermore, we have the approximation
\begin{equation}\label{HatC}
 \hat C^\nu_\alpha =  \sum_{i_{\alpha-1}, \, i_{\alpha+1} = 1}^N  \, \hat F^\nu_{\vec p, \alpha}
 \qquad (\text{with } i_\alpha \text{ fixed})
\end{equation}
for the Chern numbers $C^\nu_\alpha$ as in Eq.~\eqref{ChernNu}.
Repeating the previous argument  (for $\hat C_{\vec p}^\nu \in \mathbb Z$)
we see that $\hat C^\nu_\alpha$ is an integer,
and that $\hat C^\nu_\alpha \to C^\nu_\alpha$ for $\delta \to 0$.
This expression for $ \hat C^\nu_\alpha $ is identical to the construction in Ref.~\cite{FHS05}.

\subsection{Justification of the algorithm}

To justify the construction of our algorithm
we first note that the value of $\hat W_3$ does not depend on the position of the branch cut of the complex logarithm in Eq.~\eqref{SmallPhi}.
Indeed, if we change one $\phi_{\vec p}^\nu \mapsto \phi_{\vec p}^\nu + 2 \pi M$ by some $M \in \mathbb Z$, the integers in Eq.~\eqref{HatW3} change as
$m_{\vec p, \alpha}^\nu \mapsto m_{\vec p, \alpha}^\nu-  M$ and
$m_{\vec p + \boldsymbol\delta_\alpha, \alpha}^\nu \mapsto m_{\vec p + \boldsymbol\delta_\alpha, \alpha}^\nu + M$.
This gives an additional contribution $M \hat C_{\vec p}^\nu$ in the sum~\eqref{HatW3}.
Now if $\hat C_{\vec p}^\nu \ne 0$ we have $M_{\vec p}^\nu \mapsto M_{\vec p}^\nu - M$
(or $M_{\vec p}^{\nu'} \mapsto M_{\vec p}^{\nu'} + M$ for the associated second index $\nu'$).
Therefore, the sum~\eqref{HatW3} remains unchanged.
As one implication, we note that it does not matter whether we determine the integers $m_{\vec p, \alpha}^\nu$ using the grid points $\vec p$, $\vec p - \boldsymbol \delta_\alpha$, as we have specified before, or using the points
$\vec p$, $\vec p + \boldsymbol \delta_\alpha$.

Furthermore, this freedom in Eq.~\eqref{SmallPhi} allows us to consider two opposite choices for the $\phi_{\vec p}^\nu$ 
(even if the evaluation of Eq.~\eqref{HatW3} does not require us to make any such choice).
On the one hand, we can choose them in such a way that,
whenever a $\hat C_{\vec p}^\nu \ne 0$,
the two associated  $\phi_{\vec p}^\nu$, $\phi_{\vec p}^{\nu'}$ belong to the same branch of the complex logarithm.
Then, all $M_{\vec p}^\nu = 0$.

On the other hand, we can choose the $\phi_{\vec p}^\nu$ in such a way that they differ by less than $\pi$ on adjacent discretization points $\vec p$, $\vec p + \boldsymbol\delta_\alpha$
in the interior of the discretization grid.
For a sufficiently fine\footnote{We can understand Eq.~\eqref{Admissible} as an admissibility condition that determines the minimum possible number of discretization points.}
discretization with 
\begin{equation}\label{Admissible}
| \Phi^\nu(\vec p + \boldsymbol\delta_\alpha) - \Phi^\nu(\vec p) | < \pi
\qquad
\text{ for } 1 \le i_1, i_2, i_3 < N
\end{equation}
this choice is equivalent to setting $\phi_{\vec p}^\nu = \Phi^\nu(\vec p)$ with the `global' bands of $U(\cdot)$ as in Sec.~\ref{sec:Global}.
Then, all $m_{\vec p, \alpha}^\nu = 0$ except for the boundary values at $i_\alpha = 1$
that account for the total change of $\Phi^\nu(\mu)$ from $\mu_\alpha=0$ to $\mu_\alpha=1$.
Explicitly, we can set 
$m_{\vec p}^\nu =  W^{\alpha, \nu}$ at $i_\alpha = 1$, with the winding number of band $\nu$ as in Eq.~\eqref{WindNu}.
To prove correctness of our algorithm we now adopt this  choice,
and then compare Eq.~\eqref{HatW3} with Eq.~\eqref{W3BigPhi}.
We assume that the discretization is admissible according to Eq.~\eqref{Admissible}.

\paragraph{First proposition}
The first term in Eq.~\eqref{HatW3}  is an integer by construction.
With the present choice, the second term can be rewritten as
$\sum_{\nu=1}^n \sum_{\alpha=1}^3 \hat C^\nu_\alpha W^{\alpha, \nu}$
with the approximate Chern numbers $\hat C^\nu_\alpha$ from Eq.~\eqref{HatC},
which also gives an integer.
This shows our first proposition: $\hat W_3$ is an integer for arbitrary (admissible) discretizations.

\paragraph{Second proposition}

Now consider the limit $\delta \to 0$.
We see immediately that the second term in Eq.~\eqref{HatW3} converges to the second term in Eq.~\eqref{W3BigPhi}.
For the first term in Eq.~\eqref{HatW3}, we note that
$\hat C_{\vec p}^\nu \ne 0$ occurs only if the cube $\mathsf C_{\vec p}$ contains a degeneracy point $\mu_d$ of $U(\cdot)$.
For two degenerate eigenvalues 
$d^\nu(\mu_d) = d^{\nu'}(\mu_d)$,
with $\partial^\alpha F^\nu_\alpha (\mu_d) = -
\partial^\alpha F^{\nu'}_\alpha (\mu_d) $
and $\big(\Phi^{\nu'}(\mu_d) - \Phi^\nu(\mu_d) \big) / 2\pi  \in \mathbb Z$,
the first term in Eq.~\eqref{W3BigPhi} gives the contribution
$(1/2\pi) \big(\partial^\alpha F^\nu_\alpha (\mu_d) \big) \big( \Phi^{\nu'}(\mu_d) - \Phi^\nu(\mu_d) \big)$.
On the other hand, the first term in Eq.~\eqref{HatW3} gives the contribution
$\hat C_{\vec p}^\nu M_{\vec p}^\nu$,
and we have $2 \pi M_{\vec p}^\nu =  \phi_{\vec p}^{\nu'} -  \phi_{\vec p}^\nu = \Phi^{\nu'}(\mu_d) -  \Phi^\nu(\mu_d)$ for $\vec p \to \mu_d$.
Both terms agree in the limit $\delta \to 0$.
This shows our second proposition:
$\hat W_3$ converges to $W_3[U]$ in the limit  $\delta \to 0$.

\medskip

In combination, the second proposition shows the correctness of the algorithm,
while the first proposition justifies our constructions as it promises rapid convergence of the algorithm already on coarse discretization grids.

\subsection{Example application: $\mathrm{SU}(2)$ maps}
\label{sec:Example}

\begin{figure}
\hspace*{\fill}
\raisebox{2ex}{
\includegraphics[width=0.3\linewidth]{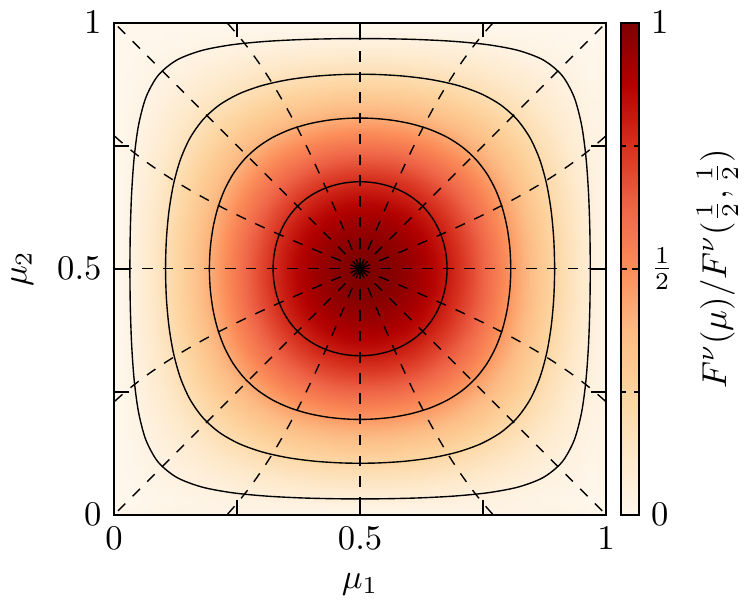}
}
\hspace*{\fill}
\includegraphics[width=0.3\linewidth]{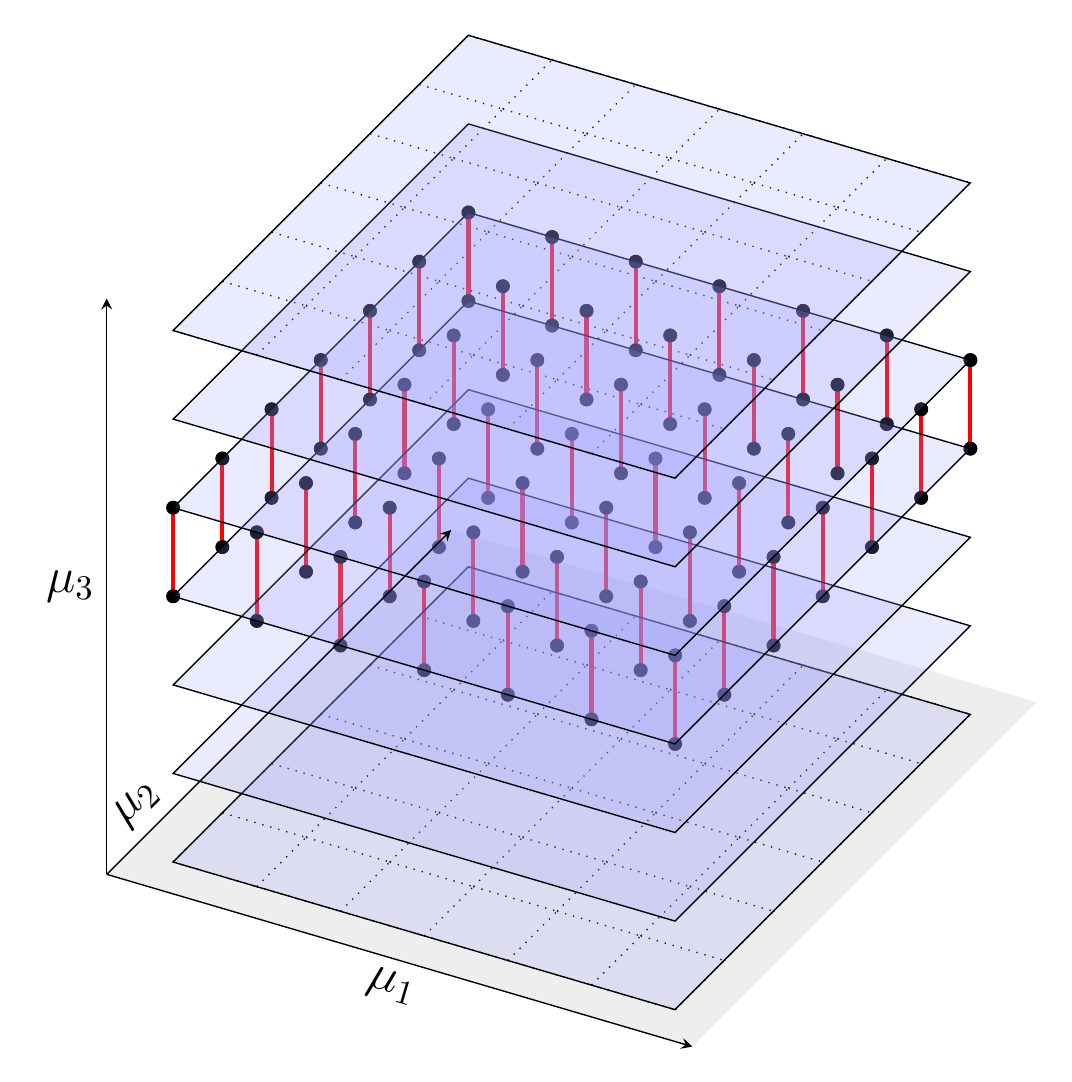}
\hspace*{\fill}
\includegraphics[width=0.3\linewidth]{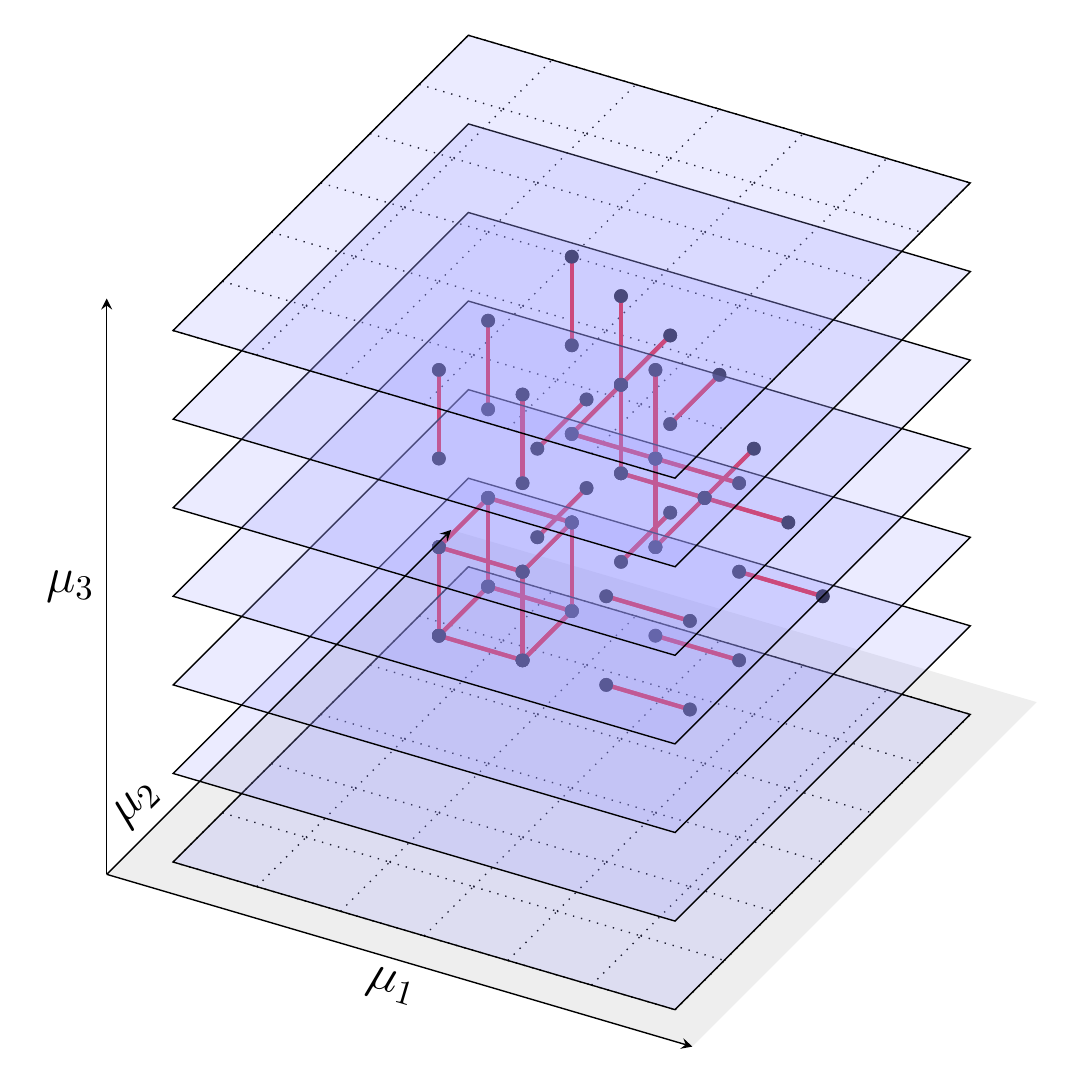}
\hspace*{\fill}
\caption{Left panel: Illustration of the map $\vec f(\cdot)$ from
the square $[0,1]^2$ to the sphere $\mathbb S^2$.
Solid (dashed) curves correspond to lines of constant $\theta$ (constant $\phi$) in spherical coordinates.
Colors give the Berry curvature $F^\nu(\mu)$ for either of the two bands of the corresponding map $U(\mu_1,\mu_2)$. 
Central and right panel: 
Position of edges $\vec p$ to $\vec p - \boldsymbol \delta_\alpha$ with a non-zero  $m_{\vec p, \alpha}^\nu$
for the first (central panel, with $w=1$) and second (right panel, with $w=2$) example map.
The opposite exterior faces in this illustration contain the same discretization points of the $6 \times 6 \times 6$ grid used here.
}
\label{fig:AlgoExample1}
\end{figure}

\begin{figure}
\hspace*{\fill}
\includegraphics[width=0.3\linewidth]{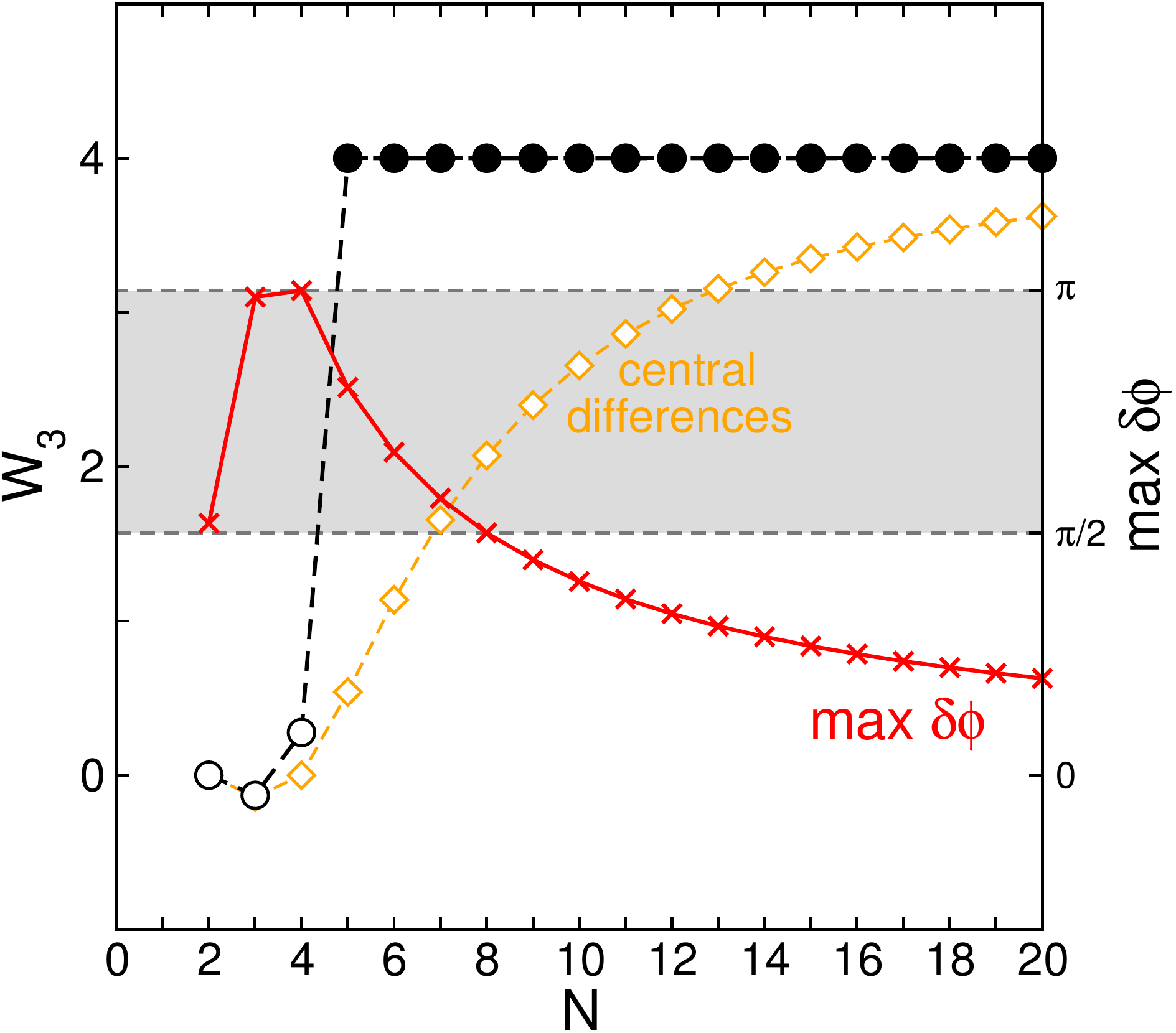}
\hspace*{\fill}
\includegraphics[width=0.3\linewidth]{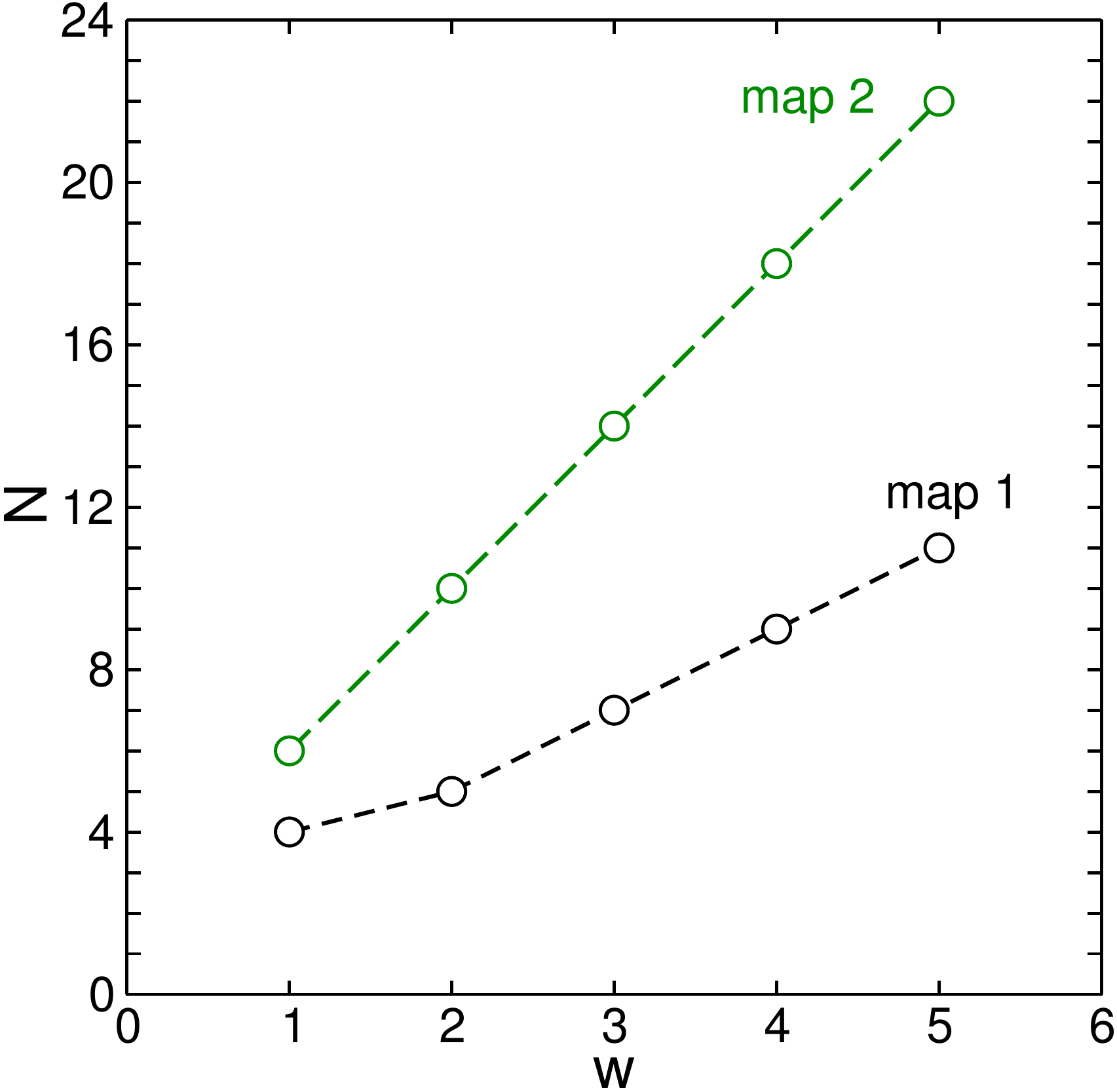}
\hspace*{\fill}
\caption{Left panel: Convergence of $\hat W_3$ for the first example map (with $w = 2$ and $W_3[U] = 4$), as a function of the number $N$ of discretization points along each direction ($N^3$ points in total).
The curve marked ``$\max \delta\phi$'' gives the maximum angle between eigenvalues in the same band on adjacent discretization points $\vec p$, $\vec p \pm \boldsymbol\delta_\alpha$. 
The curve marked ``central differences'' shows the results that would be obtained through direct evaluation of Eq.~\eqref{W3} with a central difference approximation of the derivatives $U^{-1} \partial^\alpha U$. 
Right panel: Required number $N$ for convergence $\hat W_3 \to W_3[U]$ for both example maps,
as a function of the parameter $w$ of each map.
}
\label{fig:AlgoExample2}
\end{figure}

To demonstrate the application of our algorithm, we consider maps $\mathbb T^3 \to \mathbb U(2)$ of 
spin-$\tfrac12$ rotations
\begin{equation}
U(\mu) = \exp \left[ \tfrac12 \ii  \, \vec a(\mu) \cdot  { \boldsymbol \sigma} \right]  \;.
\end{equation}
Here, $\boldsymbol \sigma$ is the vector of Pauli matrices,
and $\vec a(\cdot)$ with $\vec a(\mu) \in \mathbb R^3$ defines a vector field on $[0,1]^3$.
To obtain a map $U(\cdot)$ defined on the torus $\mathbb T^3$, suitable boundary conditions  must be fulfilled by $\vec a(\cdot)$.
In this context, recall that $U(\mu) = (-1)^n \, \mathbbm 1$ is a multiple of the identity 
if $|\vec a(\mu)| / 2 \pi \in \mathbb N_0$, independent of the direction of $\vec a(\mu)$.
We consider two examples.

\paragraph{First example}
For the first example, we 
choose a map $\vec f(\cdot)$ from the square $[0,1]^2$ to the unit sphere $\mathbb S^2 \subset \mathbb R^3$
that maps the boundary of the square to the south pole of the sphere,
and the interior of the square bijectively onto the punctured sphere without the south pole.
Then, we set $\vec a(\mu) = 4 \pi w \, \mu_3 \, \vec f(\mu_1,\mu_2)$.
Details of the map $\vec f(\cdot)$ do not matter for the value of the topological invariant $W_3[U]$.
For this map, the Berry curvature of either band of $U(\cdot)$ is a function of $\mu_1, \mu_2$,
as illustrated in Fig.~\ref{fig:AlgoExample1},
while the eigenvalues are a function of $\mu_3$.
In Eq.~\eqref{W3FY} the three-dimensional integral of $F_\alpha Y^\alpha$ thus separates.
Integration of the Berry curvature $F_\alpha$ gives the Chern number $\pm 1$ for either band,
and we have $W_3[U] = 2 w$.

\paragraph{Second example}
For the second example, 
we choose a bijective map $\vec g(\cdot)$ from the cube $[0,1]^3$ to the unit ball $|\vec r| \le 1$
that maps the surface (center) of the cube to the surface (center) of the unit ball.
Then, we set $\vec a(\mu) =2 \pi w \, \vec g(\mu)$.
The corresponding map $U(\cdot)$ has a degeneracy point at the center of the cube.
Additional degeneracies occur on concentric shells where $w | \vec g(\mu) | \in \mathbb N$.
Every additional degeneracy changes the $W_3$ invariant by $\pm 1$, and we have $W_3[U] = w$.

\medskip

We now apply our algorithm to both maps.
If the values $\phi_{\vec p}^\nu$ are computed with the standard complex logarithm,
non-zero integers $m_{\vec p, \alpha}^\nu$ contribute in the sum~\eqref{HatW3}
at the positions shown in Fig.~\ref{fig:AlgoExample1}.
For the first map, whose eigenvalues of $U(\cdot)$ change as a function of $\mu_3$,
non-zero $m_{\vec p, \alpha}^\nu$ occur in a sheet of the cube
where the eigenvalues pass through $e^{\ii \pi}$ as they move around the circle $\mathbb S^1$.
For the second map, the eigenvalues depend on the radial distance to the center and a more complicated pattern appears.
The cube around the central degeneracy point of $U(\cdot)$, where $\hat C_{\vec p}^\nu \ne 0$,
is visible in this pattern.
As explained before in connection with Eq.~\eqref{HatW3}, any other computation of 
$\phi_{\vec p}^\nu$ would result in a different arrangement of the non-zero $m_{\vec p, \alpha}^\nu$ but produce identical results for $\hat W_3[U]$.

The fast convergence of our algorithm is illustrated in Fig.~\ref{fig:AlgoExample2},
where it is compared to direct evaluation of Eq.~\eqref{W3}.
The latter, with approximation of the derivatives $U^{-1} \partial^\alpha U$ of $U(\cdot)$ by central differences\footnote{As in $f'(x) \approx (f(x+\delta) - f(x-\delta))/(2\delta)$.},
converges much more slowly.
A simple convergence criterion for our algorithm is to check that the angle between eigenvalues of $U(\cdot)$ on adjacent discretization points $\vec p$, $\vec p \pm \boldsymbol\delta_\alpha$ stays below a certain threshold,
say, below $\pi/2 = 90^\circ$.
This criterion is modelled after the admissibility condition~\eqref{Admissible}.
It guarantees that eigenvalues do not move around the entire circle $\mathbb S^1$ between adjacent discretization points.
As the data in Fig.~\ref{fig:AlgoExample2} show, correct results are already obtained with even coarser grids.
Fig.~\ref{fig:AlgoExample2} also shows the required number $N$ to achieve convergence
on an $N\times N \times N$ grid.
For the example maps, larger $N$ are required with increasing $w$ as the eigenvalues of $U(\cdot)$ change more quickly as a function of $\mu$.
In accordance with the above criterion, the required $N$ grows approximately linearly with $w$.

\section{The $W_3$ invariant for Floquet-Bloch systems}
\label{sec:Floquet}

For the application of the $W_3$ invariant to Floquet-Bloch systems,
as introduced in Ref.~\cite{PhysRevX.3.031005},
we start with the time-dependent Hamiltonian $H(t)$ of a two-dimensional lattice system. 
With translational symmetry in the $x$, $y$ lattice direction,
the Hamiltonian can be parametrized by the $k_x$, $k_y$ momentum as
$H(k_x, k_y, t)$.
Solution of the Schr\"odinger equation, either by direct integration~\cite{AF11} or 
for periodic time-dependence within the Floquet formalism~\cite{Hae97}, gives the propagator $U(k_x, k_y, t)$, with
\begin{equation}\label{UofT}
 \ii \partial_t U(k_x, k_y,t) = H(k_x, k_y, t) U(k_x, k_y,t) \;.
\end{equation}
The unitary map $U(k_x, k_y, t)$ is periodic in the $k_x$, $k_y$ direction, but not in the $t$ direction. 
For such $2+1$ dimensional maps, a construction suggested in Ref.~\cite{PhysRevX.3.031005} provides a link between the $W_3$ invariant and the number of edge states in Floquet topological insulators.
We start with the definition of the respective invariant $W_3[U_\xi]$, 
and then explain its computation with the algorithm from Sec.~\ref{sec:Algo}, which requires only a small modification of the previous scheme.
The following constructions apply to arbitrary time-dependent Hamiltonians,
although applications will often concern Floquet-Bloch systems
with a periodic Hamiltonian $H(t+T) = H(t)$.
Note that in general the Brillouin zone is not a square, but we can always assume a mapping of the crystal quasimomentum $(k_x, k_y)$ in the unit cell of the reciprocal lattice
to the coordinates $(\mu_1, \mu_2) \in [0,1]^2$, and also a mapping from time $t \in [0,T]$ to the coordinate $\mu_3 \in [0,1]$. Therefore, we will use either set of coordinates wherever appropriate.
As before, the topological invariants do not depend on details of the mapping.

\subsection{The invariant $W_3[U_\xi]$}

\begin{figure}
\hspace*{\fill}
\includegraphics[width=0.35\linewidth]{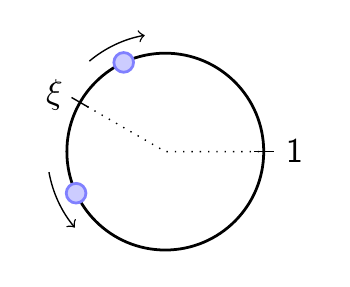}
\hspace*{\fill}
\includegraphics[width=0.35\linewidth]{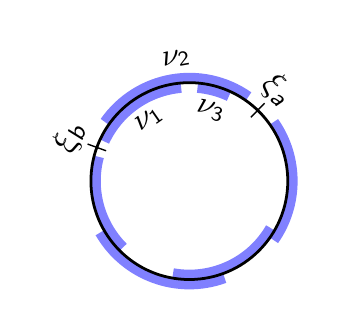}
\hspace*{\fill}
\caption{Left panel: Illustration of the action of the map $V_\xi(\cdot)$ on eigenvalues located on $\mathbb S^1$.
Right panel: Illustration of a situation where the bands of $U(\mu_1, \mu_2,1)$, indicated by thick arcs on $\mathbb S^1$, are separated by two gaps at $\xi_a$, $\xi_b$.
Here, $W_3[U_{\xi_b}] = W_3[U_{\xi_a}] - C^1_3 - C^2_3 - C^3_3$.
}
\label{fig:SketchV}
\end{figure}

The invariant $W_3[U_\xi]$ is defined for unitary maps $U(\cdot) : \mathsf [0,1]^3 \to \mathbb U(n)$
that are periodic in the $\alpha = 1,2$ directions, 
as in the first two equalities in Eq.~\eqref{BoundaryTorus},
and fulfill $U(\mu_1, \mu_2, 0) = \mathbbm 1$.
For such a map, choose a $\xi \in \mathbb S^1$ that differs from all eigenvalues $d^\nu(\mu_1,\mu_2,1)$ 
at $\mu_3=1$, i.e.,
that lies in a gap between the bands $(\mu_1,\mu_2) \mapsto d^\nu(\mu_1,\mu_2,1)$ of $U(\cdot, \mu_3 = 1)$. For time-periodic systems, these bands are often called the Floquet bands.
Note that the existence of a gap implies that the winding numbers $W^{1,\nu}$, $W^{2,\nu}$  of every such band are zero, which agrees with the assumption $U(\mu_1, \mu_2, 0) = \mathbbm 1$.

In this situation, we can consider the map
\begin{equation}\label{VMap}
 V_\xi(s,z) = \exp\big( (1-s) \log_\xi z  \big) \;,
\end{equation}
where $\log_\xi z$ is the complex logarithm with branch cut along the line 
from zero through $\xi$.
As $s=0$ increases to $s=1$, the map $s \mapsto V_\xi(s,e^{\ii \phi})$ rotates eigenvalues $e^{\ii \phi}$ on the two different circle segments from one to $\xi = e^{\ii \epsilon}$ clockwise (as $s \mapsto \exp(\ii (1-s) \phi)$ for $0 \le \phi < \epsilon \le 2\pi$)
or counterclockwise (as $s \mapsto \exp( \ii  (1-s) \phi + 2 \pi \ii s )$ for $0 \le \epsilon < \phi \le 2\pi$)
to one
(see Fig.~\ref{fig:SketchV}).
The invariant $W_3[U_\xi]$
is now computed 
for the map 
\begin{equation}\label{UXi}
 U_\xi(\mu) = \begin{cases}
 U(\mu_1, \mu_2, 2 \mu_3) \quad & \text{ if } 0  \le \mu_3 \le 1/2 \;, \\
 V_\xi \big[2\mu_3-1 , U(\mu_1, \mu_2, 1) \big]   & \text { if } 1/2 \le \mu_3 \le 1 \;,
  \end{cases}
\end{equation}
which is periodic in all directions $\alpha=1,2,3$.

In Eq.~\eqref{W3FY} the contribution from the $V_\xi(\cdot)$ part of Eq.~\eqref{UXi} can be evaluated explicitly, and we have
\begin{equation}\label{W3Xi}
 W_3[U_\xi] = W_3[U] - \frac{1}{2 \pi \ii} \int\limits_0^1 \!\!\! \int\limits_0^1  \tr \big[ F_3( \mu_1, \mu_2, 1 )  \log_\xi[ D(\mu_1, \mu_2,1)]  \big] \,  \mathrm d\mu_1 \mathrm d\mu_2 \;,
\end{equation}
where the value of $W_3[U]$ is obtained by evaluation of Eq.~\eqref{W3FY} for the map $U(\cdot)$ (which by itself does not give a topological invariant because $U(\cdot)$ is not periodic in the $\alpha=3$ direction).
Note that the $\xi$-dependent correction term in this expression depends only on
the Berry curvature and eigenvalues of $U(\cdot)$ at $\mu_3=1$.

When the value $\xi$ 
changes from a gap at $\xi = \xi_a$ to a gap at $\xi = \xi_b$
it will pass through some bands $\nu_1, \nu_2, \dots, \nu_k$ of $U(\cdot, \mu_3=1)$ (see Fig.~\ref{fig:SketchV}).
For eigenvalues in these bands, the logarithms $\log_\xi d^\nu(\mu_1,\mu_2,1)$ in Eq.~\eqref{W3Xi} change by $\pm 2\pi \ii$ (for example, $\log_{\xi_b} d^\nu =\log_{\xi_a} d^\nu + 2 \pi \ii$ for  $\nu = 1,2,3$ in Fig.~\ref{fig:SketchV}).
From comparison with Eq.~\eqref{ChernNu} we see that the entire expression changes by the Chern numbers 
$C^\nu_3$ of these bands. We thus get the relation
\begin{equation}\label{W3XiaXib}
 W_3[U_{\xi_b}] = W_3[U_{\xi_a}] - \sum_{\nu = \nu_1, \dots, \nu_k} C^\nu_3 \big|_{\mu_3 = 1}
\end{equation}
between the $W_3[U_\xi]$ invariant in two gaps $\xi_a$, $\xi_b$, and the Chern numbers of the bands lying on the counterclockwise circle segment from $\xi_a$ to $\xi_b$.
If $\xi$ moves once around the entire circle, we recover\footnote{This observation also resolves the potential ambiguity that occurs in Eq.~\eqref{VMap} for $\xi \to 1$.} the identity $W_3[U_\xi] = W_3[U_\xi]$
since the Chern numbers add up to $\sum_{\nu=1}^n C^\nu_\alpha = 0$.

\subsection{Computation of $W_3[U_\xi]$}
\label{sec:AlgoW3Xi}

We can immediately apply the algorithmic scheme of Sec.~\ref{sec:Algo} to the computation of $W_3[U_\xi]$
if we absorb the $\xi$-dependent term in Eq.~\eqref{W3Xi} into a correction to Eq.~\eqref{HatW3}.
This correction results only from the difference of $\phi_{\vec p}^\nu$ and $-\ii \log_\xi  d^\nu(\vec p)$ at $i_3 = N$.

First, we apply the previous algorithm to the map $U(\cdot)$ and compute $\hat W_3$ from the sum~\eqref{HatW3} (which by itself does not give an integer because $U(\cdot)$ is not periodic).
Because of $U_\xi(\cdot) = \mathbbm 1$ for $\mu_3 = 0,1$,
we can set $m^\nu_{\vec p, \alpha} = M^\nu_{\vec p}=0$ for $i_3 = 1$, 
such that the sum starts effectively with $i_3=2$,
and $M^\nu_{\vec p}=0$ for $i_3 = N$.
Second, to account for the $\xi$-dependent correction, we determine integers $K_{i_1, i_2}^\nu$ such that
 $|-\ii \log_\xi d^\nu(\vec p) - \phi_{\vec p}^\nu  + 2 \pi K_{i_1, i_2}^\nu| < \pi$
 at $i_3 = N$.
Then, we have
\begin{equation}\label{HatW3Xi}
 \hat W_3[U_\xi] = \hat W_3 + \sum_{i_1, i_2 = 1}^N \sum_{\nu=1}^n  K_{i_1, i_2}^\nu \,  \hat F_{\vec p, 3}^\nu \Big|_{i_3=N}  \;.
 \end{equation}
As before, $\hat W_3[U_\xi]$ is an integer and converges to $W_3[U_\xi]$ for $\delta \to 0$.
In this expression, only the second term depends on $\xi$,
and reuses quantities already computed in the algorithm.

Note that we can repeat the argument leading to Eq.~\eqref{W3XiaXib}:
If $\xi$ moves from gap $\xi = \xi_a$ to another gap $\xi=\xi_b$ the integers in Eq.~\eqref{HatW3Xi} change by $K_{i_1, i_2}^\nu \mapsto K_{i_1, i_2}^\nu - 1$ for bands $\nu = 1, \dots, k$ lying between the two gaps.
According to Eq.~\eqref{ChernNu}, 
the value of $\hat W_3[U_\xi]$ thus changes by the approximate Chern numbers $-(\hat C^1_3 + \cdots +  \hat C^k_3)$ of the respective bands, 
which agrees with Eq.~\eqref{W3XiaXib}.
The Chern numbers $\hat C^\nu_3$ are also provided by the algorithm.

\subsection{Tracking the evolution of $W_3[U_\xi]$ for gapped systems}
\label{sec:Track}

\begin{figure}
\hspace*{\fill}
\includegraphics[width=0.35\linewidth]{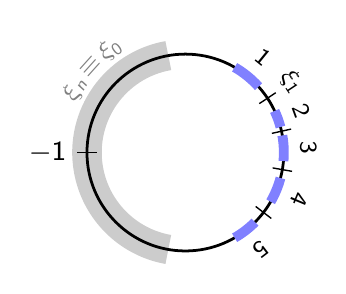}
\hspace*{\fill}
\includegraphics[width=0.35\linewidth]{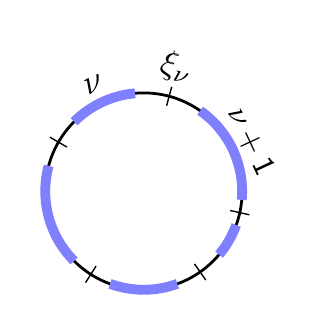}
\hspace*{\fill}
\caption{Sketch of typical configurations of the bands of $U(\cdot, \mu_3)$, indicated by thick arcs on $\mathbb S^1$.
At small $\mu_3$ (left panel), all bands lie in the vicinity of the point $1 \in \mathbb S^1$, before they can wrap around the circle for larger $\mu_3$ (right panel).
Bands are numbered clockwise, and the gap at $\xi_\nu$ separates bands $\nu$ and $\nu+1$ .
For small $\mu_3$, the $n$-th gap $\xi_n \equiv \xi_0$ can be chosen anywhere around $- 1 = e^{\pm  \ii \pi}$.
}
\label{fig:SketchBands}
\end{figure}

The invariant $W_3[U_\xi]$ can be considered as a function of time $t$,
and in physics applications we might want to track the change of $W_3[U_\xi]$ with increasing $t$.
Note that $W_3[U_\xi]$ 
depends on the entire history of $U(k_x, k_y, t')$ for $0 \le t' \le t$,
not only on the map $U(k_x, k_y, t)$ at the final time $t$.

To describe how the change of $W_3[U_\xi]$  can be tracked 
we assume that,
apart from isolated degeneracy points, the system is fully gapped for all $t>0$.
Then, the $n$ bands $\nu = 1, \dots, n$ of $U(\cdot, t)$ are separated by $n$ gaps at positions $\xi_1, \dots, \xi_n$,
as illustrated in Fig.~\ref{fig:SketchBands}.
Because the relation $d^\nu \equiv \exp(-\ii \epsilon^\nu t)$
between eigenvalues $d^\nu$ of $U(\cdot,t)$  and (Floquet quasi--) energies $\epsilon^\nu$ involves a minus sign, we number bands and gaps clockwise on $\mathbb S^1$.
If we count the index $\nu$ modulo $n$, we have $\xi_n = \xi_0$.

To each gap, we assign the value $n^\nu = W_3[U_{\xi_\nu}]$.
It does not depend on the precise location of $\xi_\nu$ within the gap,
but only on the gap index $\nu$.
According to the bulk-edge correspondence for Floquet-Bloch systems, $n^\nu$ gives the number of edge states in the respective gap~\cite{PhysRevX.3.031005}.

 According to Eq.~\eqref{W3XiaXib},
the different $n^\nu$ are related to the Chern numbers of the (Floquet) bands via
\begin{equation}
\label{nandc}
n^{\nu+1} = n^\nu + C^{\nu+1}_3 \;.
\end{equation}
For small times $t$, before the bands can wrap around the circle $\mathbb S^1$,
we have $W_3[U_{\xi_n}] = 0$ and thus
\begin{equation}
\label{nfromc}
n^\nu = C^1_3 + \dots + C^{\nu}_3   \qquad\qquad (\text{if } W_3[U_{\xi_n}] = 0) \;.
\end{equation}
Especially, we have $C^1_3 + \dots + C^n_3 = 0$ in accordance with $n^n = n^0 = 0$.

As the systems evolves in time,
gaps close and reopen at isolated~\cite{Wigner} degeneracy points of $U(\cdot)$.
Computing the precise location of these points is impractical. Instead, we want to propagate the system over an entire step $t \mapsto t + \delta t$ (or $\mu_3 \mapsto \mu_3 + \delta$ in $\mu$ coordinates), and 
determine the values $\tilde n^\nu$ at $t + \delta t$ from the values $n^\nu$ at $t$.

Within our algorithm, performing a time step 
corresponds to extending the discretization grid from $i_3 = N$ to $i_3=N+1$.
How the values $n^\nu$ change can now be deduced from Eq.~\eqref{HatW3Xi}.
In the special situation of a fully gapped system considered here,
where the same gap $\nu$ closes and reopens at a degeneracy point,
all the integers $K_{i_1, i_2}^\nu$ stay the same in this equation\footnote{To be precise, they stay the same if we set $\phi^\nu_{\vec p} = - \ii \log_\xi d^\nu(\vec p)$.
As noted earlier in connection with Eq.~\eqref{HatW3}, the choice of the branch cut of the complex logarithm does not affect the algorithm.}.
The value $n^\nu$ changes only if,
by extending the grid, we capture a new degeneracy point of $U(\cdot)$ that gives a non-zero $M^\nu_{\vec p} \hat C_{\vec p}^\nu = M^{\nu+1}_{\vec p}  \hat C_{\vec p}^{\nu+1}$ contribution at $i_3 = N$ for the bands $\nu$, $\nu+1$ next to the gap~\cite{1367-2630-17-12-125014}.
Recalling the definition of $M^\nu_{\vec p}$ in Eq.~\eqref{HatW3} and of $\log_\xi$ in Eq.~\eqref{HatW3Xi}, we find that $M^\nu_{\vec p}= - M^{\nu+1}_{\vec p} = 1$ 
if $\xi = \xi_\nu$ lies in the gap between bands $\nu$, $\nu+1$,
and  $M^\nu_{\vec p}=  M^{\nu+1}_{\vec p} =0$ otherwise. 

Therefore, if the algorithm detects a non-zero $\hat C_{\vec p}^\nu = - \hat C_{\vec p}^{\nu+1} \ne 0$ at the present $i_3 = N$, a degeneracy point in gap $\nu$ occurs between $t$ and $t + \delta t$.
The values $n^\nu$ change as
\begin{equation}\label{TrackN}
\tilde n^\nu = n^\nu + \hat C_{\vec p}^\nu  = n^\nu - \hat C_{\vec p}^{\nu+1}   \;,
\qquad \text{ and }
\quad
\tilde n^{\nu'} = n^{\nu'}  \quad \text{ for } \nu' \ne \nu \;.
\end{equation}
At the same time, the Chern numbers of the two bands next to the gap change as
$\tilde C^\nu_3 = C^\nu_3 + \hat C_{\vec p}^\nu$, 
$\tilde C^{\nu+1}_3 = C^{\nu+1}_3 - \hat C_{\vec p}^\nu$.
This replacement leaves the relation~\eqref{nandc} intact.
If more than two non-zero $\hat C_{\vec p}^\nu$ occur in the same slice, they have to be summed.

The above scheme to track the change of the values $n^\nu$ directly is a certain simplification of the more general algorithm from Sec.~\ref{sec:AlgoW3Xi}.
We should note, however, that the tracking scheme applies primarily to fully gapped systems as in Fig.~\ref{fig:SketchBands},
while the general algorithm is preferable in situations as in Fig.~\ref{fig:SketchV} where bands overlap during time evolution.
Note that in the practical computation, the cubes with non-zero $\hat C_{\vec p}^\nu$ that enter in Eq.~\eqref{TrackN} are simply obtained with Eq.~\eqref{Cpnu} evaluated on the discretization grid, as explained previously for our algorithm in Sec.~\ref{sec:Algo}.
In particular, we do not need to determine the precise positions of the degeneracy points of $U(\cdot)$.

\subsection{Specialization to static systems}

Let us briefly note how the invariant $W_3[U_\xi]$ generalizes the bulk-edge correspondence for static systems, which involves only the Chern numbers of the respective bands,
and how this fact manifests itself in our algorithm.

For static systems with a constant Hamiltonian $H(k_x,k_y,t) \equiv H(k_x,k_y)$,
the matrix $S(\cdot)$ in the decomposition~\eqref{USD} does not depend on $t \equiv \mu_3$,
while the eigenvalues $d^\nu(\mu) = \exp(-\ii \mu_3 E^\nu(\mu_1,\mu_2) )$
change at a constant rate given by the eigenvalues $E^\nu(\mu_1,\mu_2) \equiv E^\nu(k_x, k_y)$ of $H(k_x, k_y)$, very much as in the first example map in Sec.~\ref{sec:Example}.

For small $t \equiv \mu_3$, we have a situation as in Fig.~\ref{fig:SketchBands} (left panel),
and according to the previous considerations the values $n^\nu = W_3[U_\xi]$ in the different gaps differ by the (constant) Chern numbers of the bands of $H(k_x,k_y)$.
In this way, we recover with Eq.~\eqref{nfromc} the bulk-edge correspondence for static systems, which relates the number of edge states $n^\nu$ and the Chern numbers $C^\nu_3$ of the bands of $H(k_x,k_y)$.
Our algorithm provides the numbers $n^\nu$ immediately, in the first step at $i_3=1$:
Once as the result of the direct computation of $W_3[U_\xi]$,
and again via summation of the approximate Chern numbers $\hat C^\nu_3$ from Eq.~\eqref{ChernNu}.

\section{Application: Anomalous edge states in irradiated graphene}
\label{sec:Graphene}

\begin{figure}
\hspace*{\fill}
\includegraphics[width=0.4\columnwidth]{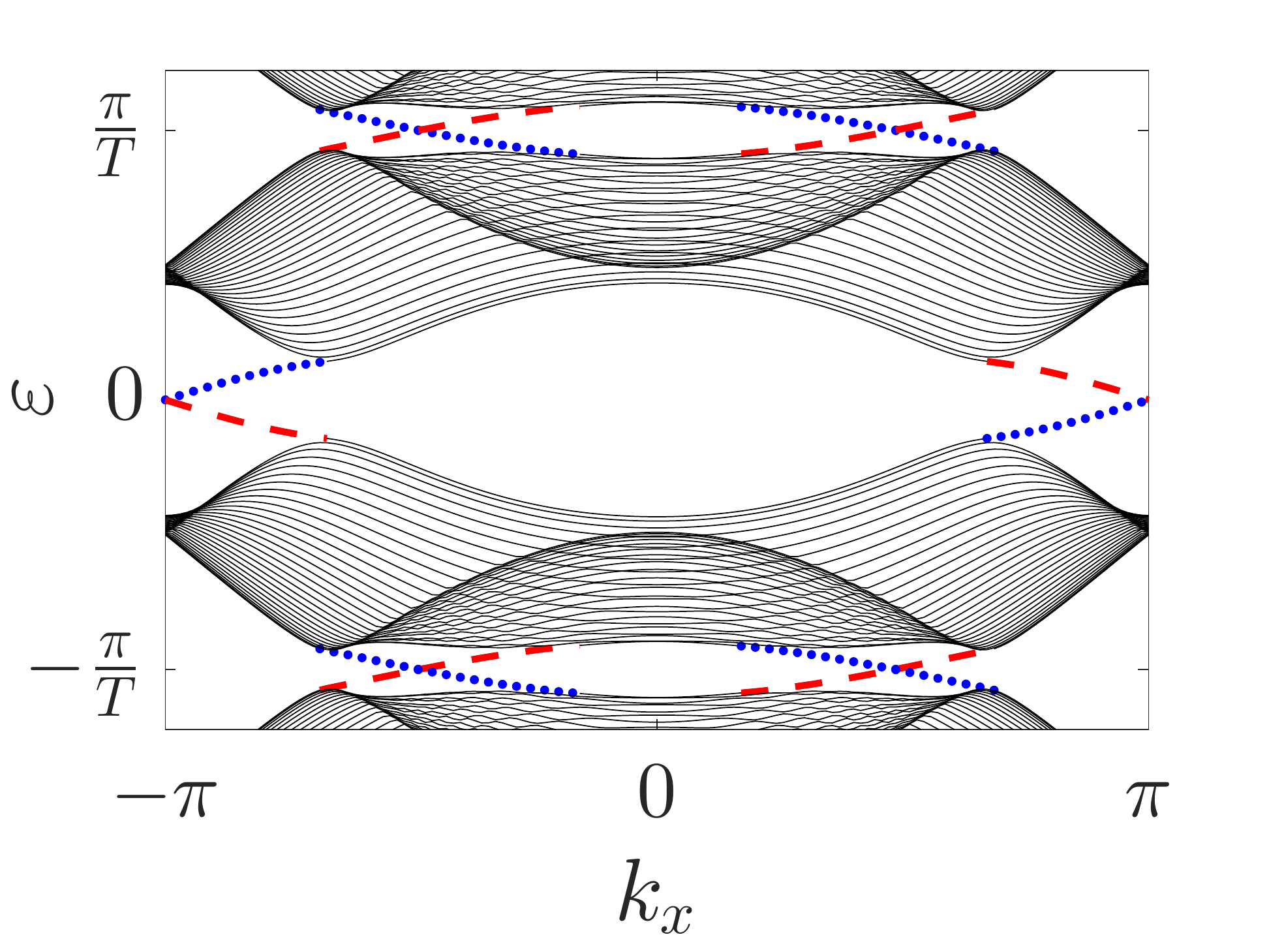}
\hspace*{\fill}
\caption{The quasienergy spectrum $\epsilon^{\nu}(k_x)$ of the Hamiltonian~\eqref{H_graphen} in a finite strip geometry with zigzag edges along the $x$ direction. Here we choose $A_0=0.7$ and $\omega=3.5$. 
Two gaps exist at quasienergies $\epsilon_1 = 0$ (i.e., $\xi_1 = 1$) and
$\epsilon_2 = \pi/T$ (i.e, $\xi_2 = -1$).
The chiral edge states in each gap, which are localized on the bottom (top) edge of the strip, are indicated by dashed red (dotted blue) curves. We count $n^1 = -1$ and $n^2 = 2$.}
\label{fig:GrapheneSpec}
\end{figure}

After introduction of our algorithm we now want to apply it to a representative Floquet-Bloch system. To this end, we consider 
the Hamilton operator~\cite{PhysRevLett.107.216601, PhysRevB.90.115423, PhysRevLett.114.056801, Wang20161} (omitting irrelevant constants)
\begin{equation}
H_\mathrm{graphene}(t)= \sum_{\langle i,j \rangle} e^{\ii A_{ij}(t)}c_i^{\dagger} c^{}_j +e^{-\ii A_{ij}(t)}c_j^{\dagger} c^{}_i 
\label{H_graphen}
\end{equation}
of a  $2$-band tight-binding model on a graphene lattice,
where irradiation with circularly polarized light 
is included through a time-periodic Peierls~\cite{Peierls} phase
$A_{ij}(t)=\mathbf{A}(t) \cdot (\mathbf{R}_i-\mathbf{R}_j)$.
Here, $\vec R_i - \vec R_j$ is the lattice vector from site $j$ to site $i$,
and $\mathbf{A}(t)=A_0\left(\sin(\omega t), \cos(\omega t)\right)$ gives the light wave with amplitude $A_0$ and frequency $\omega$.
Without driving ($A_0=0$) the Hamiltonian~\eqref{H_graphen} reduces to the standard graphene tight binding model, with linear electron dispersion at the Dirac points. There are no chiral edge states and the system is topologically trivial.

Irradiation with light ($A_0 \ne 0$) induces a band gap at the Dirac points, and chiral edge states appear in this gap. Through variation of the light amplitude and frequency various topological phase transitions can be observed~\cite{Wang20161}. 
In remarkable contrast to the static case,
where additional coupling terms such as spin orbit coupling are needed in the Hamiltonian to induce
the transition from a topologically trivial to a nontrivial state,
simple time-dependent variation of the Peierls phase suffices to establish a Floquet topological insulator.

Here, we consider a parameter regime ($A_0=0.7$ and $\omega=3.5$) where anomalous edge states appear in the Floquet band structure (see Fig.~\ref{fig:GrapheneSpec}). 
The two Floquet bands, between the gaps at $\xi_1 =1$ and $\xi_2 = -1$, have Chern numbers $C_3^1=-3$ and $C_3^2=3$. For static systems, we would expect from the bulk-edge correspondence~\eqref{nfromc} that three chiral edge states exist in the band gap between the two bands. However, Floquet quasienergies are defined only up to multiples of $2\pi/T$ and no ``lowest'' or ``highest'' band exists, such that it is no longer possible to predict the number of edge states from the Chern numbers alone.
Indeed, Fig.~\ref{fig:GrapheneSpec} shows that the correct net-chirality of edge states in the two band gaps  is $n^1=-1$ and $n^2=2$. The relation~\eqref{nandc} holds instead of the simpler Eq.~\eqref{nfromc}.
 Therefore, computation of the winding number $W_3[U_{\xi}]$ is essential for the topological classification of such a Floquet-Bloch system.

\begin{figure}
\hspace*{\fill}
\includegraphics[width=0.38\linewidth]{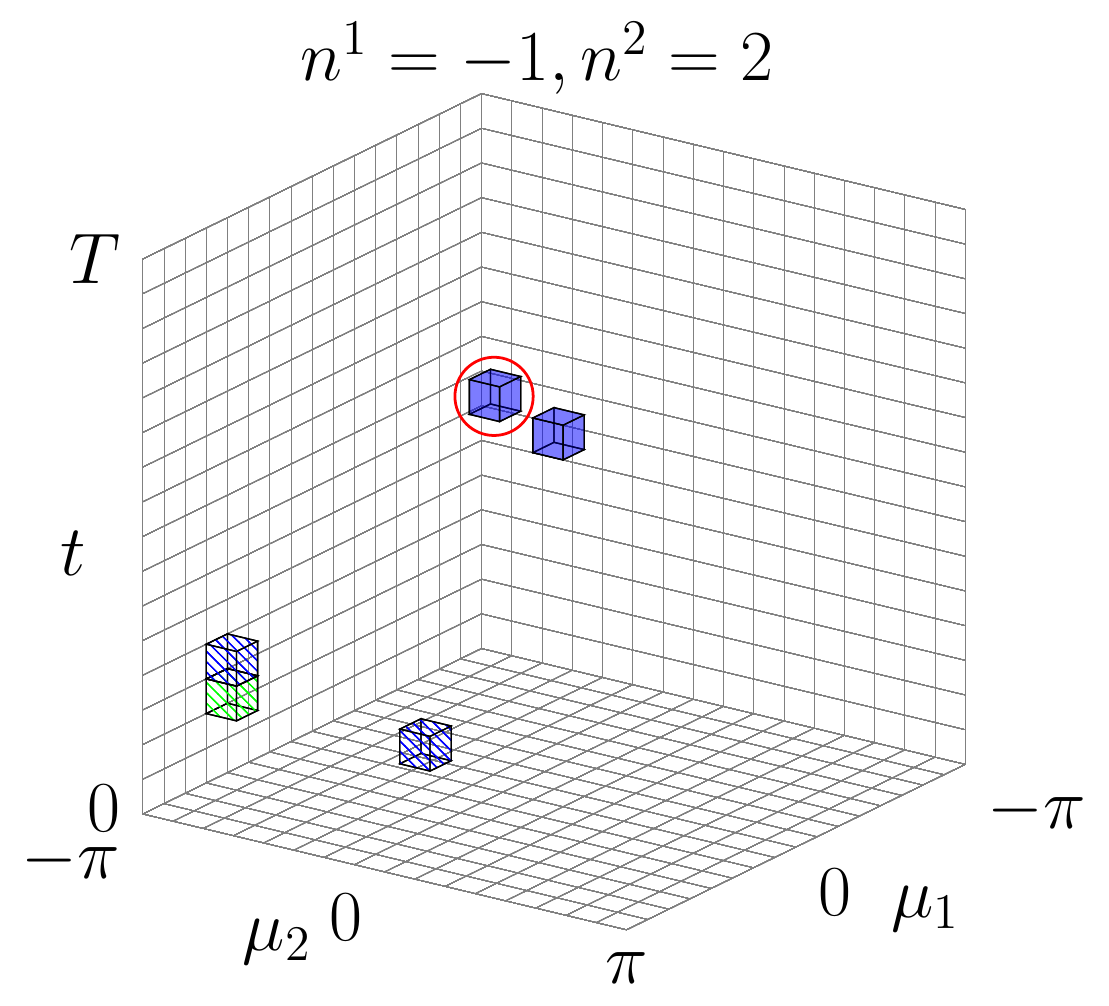}
\hspace*{\fill}
\includegraphics[width=0.38\linewidth]{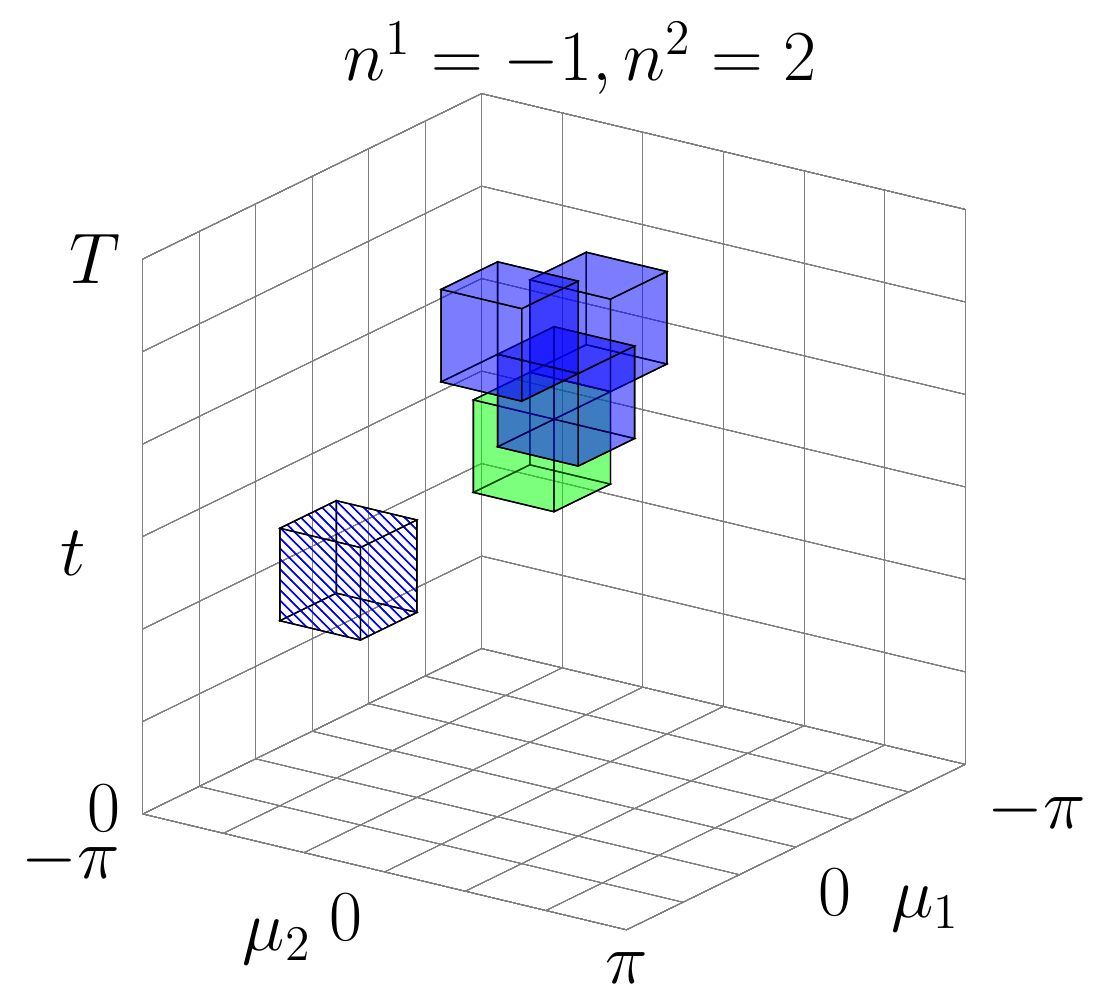}
\hspace*{\fill} 
\caption{Position of cubes $\mathsf C_{\vec p}$ with non-zero Chern numbers $\hat C^\nu_{\vec p}$ for $A_0=0.7$ and $\omega=3.5$,
on  a fine ($16\times 16\times 16$, left panel) and coarse ($6\times 6\times 6$, right panel) discretization grid.
Hatched (filled) cubes change the value of $n^1$ ($n^2$) in the band gap at $\xi_1$ ($\xi_2$) during execution of the algorithm according to Eq.~\eqref{TrackN},
with $\hat C^{1}_{\vec p}=1$ ($\hat C^{1}_{\vec p}=-1$) for green (blue) cubes.
 The values $n^{\nu}$ above each panel are the results obtained for the respective grid.
 Both grids give the correct converged result.
}
 \label{fig:GrapheneCubes}
\end{figure}

\begin{figure}
\hspace*{\fill}
\includegraphics[width=0.32\linewidth]{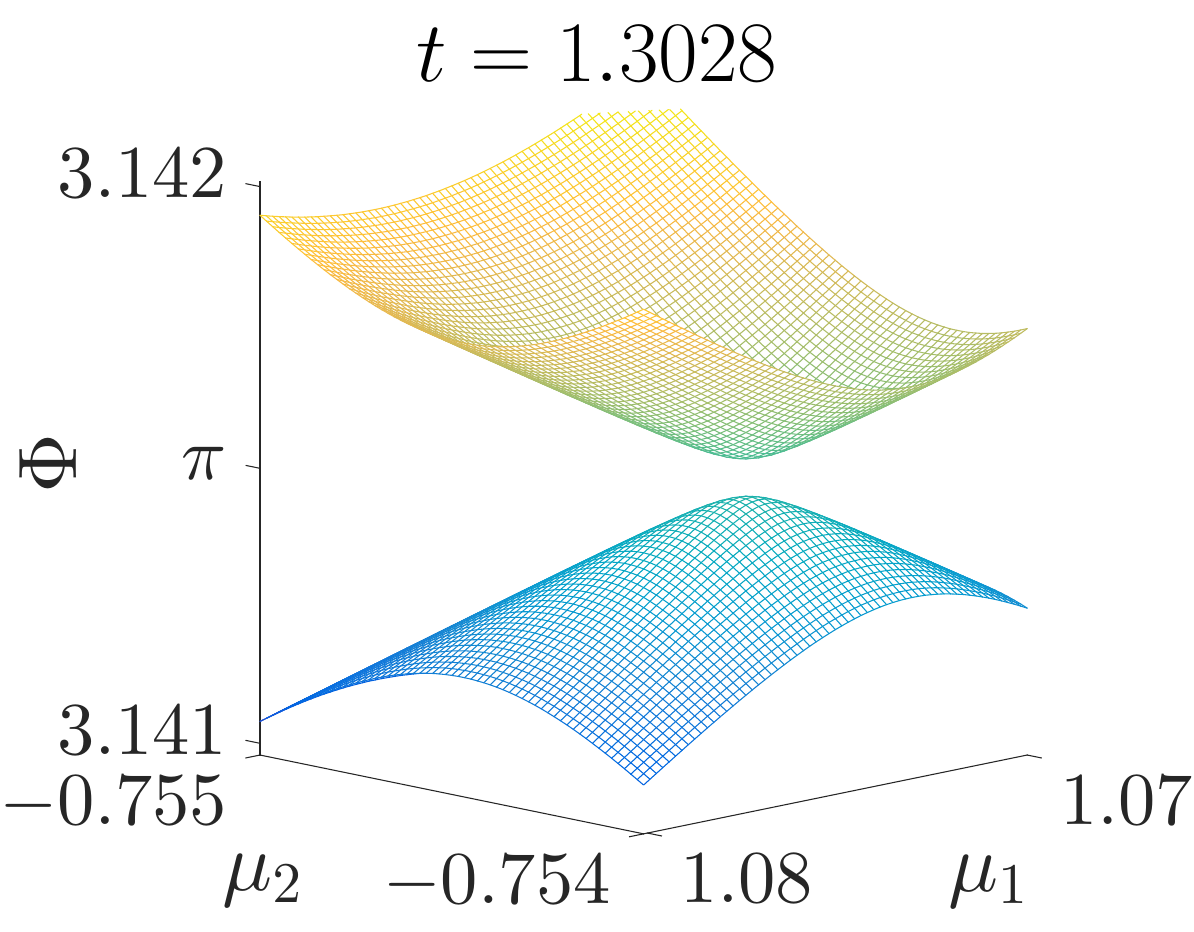}
\hspace*{\fill}
\includegraphics[width=0.32\linewidth]{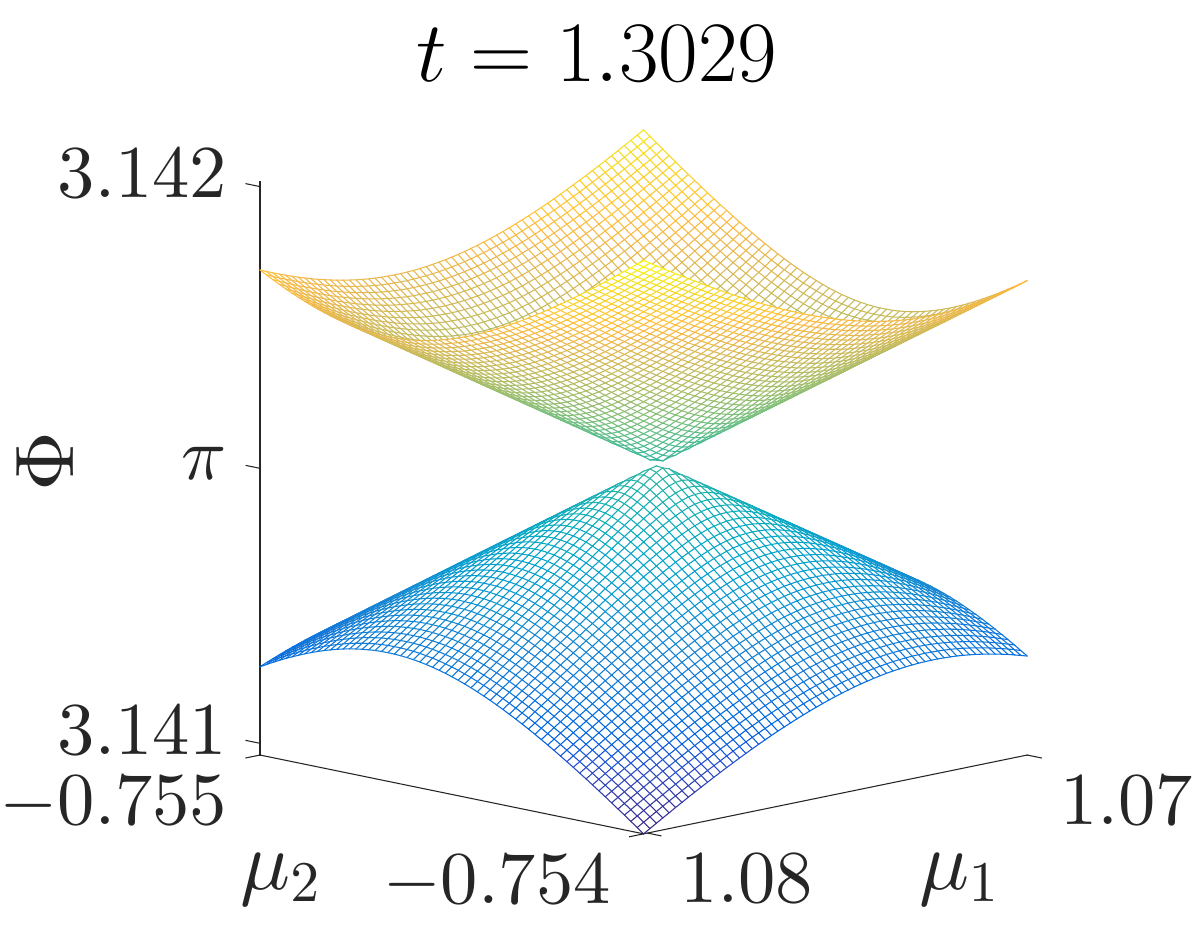}
\hspace*{\fill}
\includegraphics[width=0.32\linewidth]{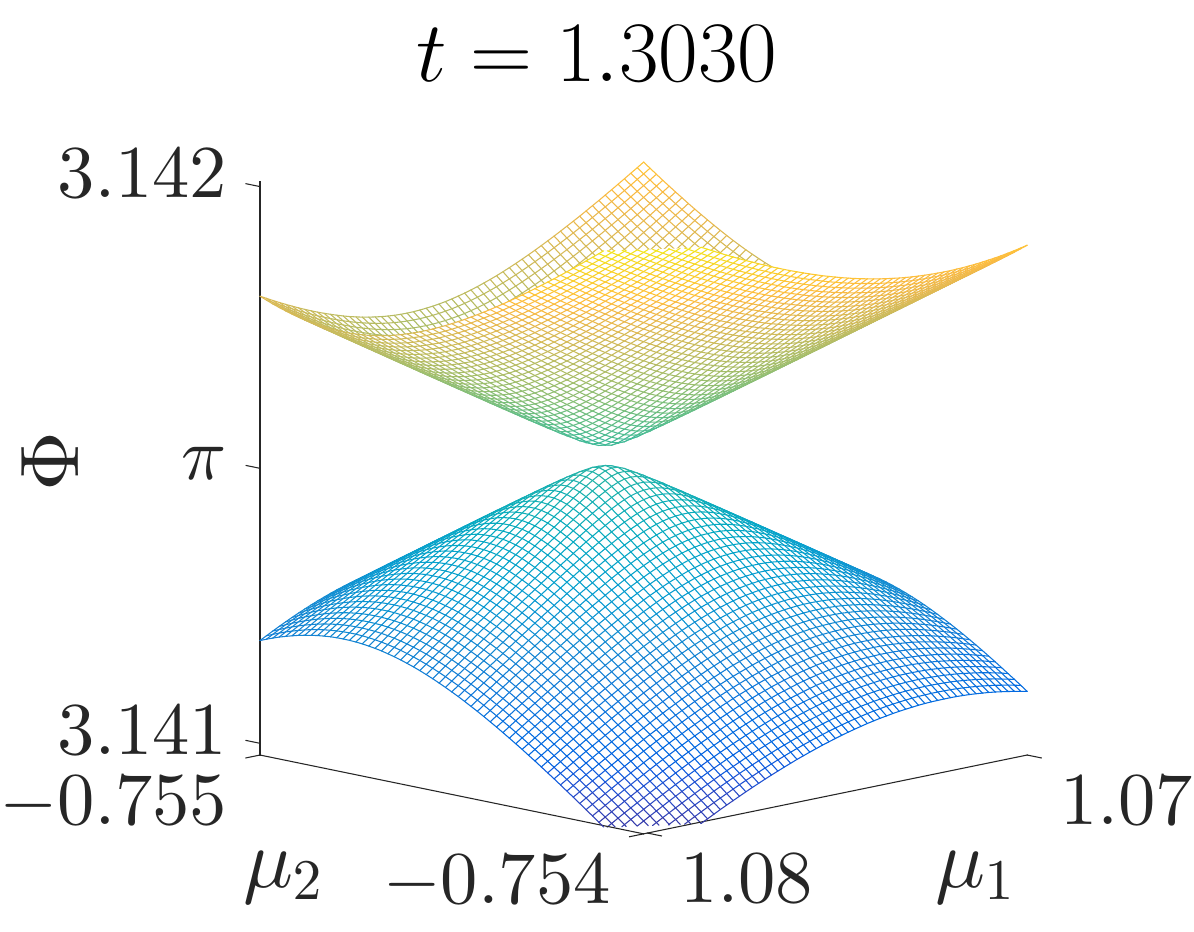}
\hspace*{\fill}
\caption{Time evolution of the bands $\Phi^{\nu}(k, t) \equiv \Phi^{\nu}(\mu_1,\mu_2, \mu_3)$
 inside the red circled cube in Fig.~\ref{fig:GrapheneCubes} (a), showing that the cube encloses a degeneracy point. To better visualize the degeneracy the coordinates 
 $(\mu_1, \mu_2)$
 were rotated by $35^{\circ}$.}
 \label{fig:GrapheneCross}
\end{figure}

We now compute $n^1$ and $n^2$ directly with the algorithm introduced in Sec.~\ref{sec:Floquet}. 
In the following, we map the crystal momentum $k_x$, $k_y$ 
to a square with coordinates $\mu_1, \mu_2$,
which is equivalent to a rhombic discretization of the hexagonal Brillouin zone of graphene.
Fig.~\ref{fig:GrapheneCubes} shows the cuboids with non-zero $\hat C^\nu_{\vec p}$ that occur in the algorithm for a fine (coarse) grid with $16 \times 16 \times 16$ ($6\times 6\times 6$)  discretization points.
As discussed previously, such cuboids contain a degeneracy point where two (Floquet) bands touch.
A close-up view of such a degeneracy point is shown in Fig.~\ref{fig:GrapheneCross}, where we zoom into the cuboid marked in Fig.~\ref{fig:GrapheneCubes} (left panel) with a red circle.
At the degeneracy point, local expansion of $U(\cdot)$ would provide the necessary information to predict how the values $W_3[U_\xi]$ and $n^\nu$ change as the gap closes,
but isolation of the degeneracy point is impractical and requires high computational effort with many evaluations of $U(\cdot)$.
Instead, comparison of the computation on the fine and coarse grid shows that the present algorithm 
produces the correct values $n^1=-1$ and $n^2=2$ already on the coarse grid.
In the present example, this grid also gives the minimal number of discretization points required to obtain correct results.
Coarser grids, for which wrong results would be produced by the algorithm,
 violate the admissibility conditions formulated in Sec.~\ref{sec:Algo}.
The simple check from Sec.~\ref{sec:Example} on the maximum angle between eigenvalues on adjacent discretization points (as in Fig.~\ref{fig:AlgoExample2}) guarantees correctness of the algorithm.

\section{Conclusions}
\label{sec:Conc}

In this paper we introduce an algorithm for the efficient computation of the $W_3$ invariant,
and demonstrate its application to general unitary maps and to the propagators of time-dependent (Floquet-) Bloch systems.
In both cases, the $W_3$ invariant is computed with moderate effort,
requiring evaluation of the unitary map only at a small number of discretization points.

The construction of the algorithm relies on an expression for the $W_3$ invariant in terms of the Berry curvature of the eigenvectors 
and the angular velocity of the eigenvalues of the unitary map.
Rapid convergence with the number of discretization points is guaranteed by the fact that the algorithm generates, indepently of the discretization, an integer number.
In this respect, our algorithm is related to the algorithm of Fukui, Hatsugai, Suzuki~\cite{FHS05},
where a similar construction is used for the computation of Chern numbers.
We note specifically that our algorithm does not attempt to simply evaluate Eq.~\eqref{W3} with some differentiation and integration rule.
Instead, it directly extracts topological information about $U(\cdot)$ that is contained in the definition of the $W_3$ invariant.
In addition to the computation of Chern numbers the algorithm for the $W_3$ invariant must also account for the winding of eigenvalues.
In this respect, our algorithm extends the algorithm from Ref.~\cite{FHS05}.

Concerning the physical applications of the algorithm,
we note that the $W_3$ invariant can be computed for arbitrary time-dependence, not only for the periodic time-dependence of Floquet-Bloch systems.
While observation of topological properties for non-periodic time-dependence may be difficult in solid state systems,
one should be able to use photonic crystals~\cite{nature12066,nphoton.2014.248} to implement such scenarios.
The present algorithm will then allow us to determine topological properties also in complicated situations where alternative methods of computation fail.

From a mathematical point of view,
one might want to note that the $W_3$ invariant in Eq.~\eqref{W3} is entirely formulated in terms of the unitary Lie group and Lie algebra, while the present algorithm explicitly refers to the matrix realization of their elements.
Out of mathematical curiosity it is natural to ask for a ``Lie-type'' algorithm that does not require spectral decomposition of the unitary map. 
Such an algorithm can be easily constructed for the $W_1$ invariant,
but we do not know of a similar algorithm for the $W_3$ (or higher dimensional) invariant
that preserves the favorable properties of the present algorithm,
e.g., by producing integer number results only.
Clearly, for the $W_3$ invariant the situation complicates because the different derivatives in Eq.~\eqref{W3} do not need to commute.
In any case, for the physical applications considered here the present formulation in terms of eigenvectors and eigenvalues is both conceptually and computationally adequate.
The interpretation that non-trivial topology is generated in time-dependent systems through band ``crossings'' at degeneracy points gives the correct picture for the physical applications.
The implementation of the present algorithm is reasonably easy,
and as application to the irradiated graphene system demonstrates the algorithm provides us with a computationally efficient solution to the problem of computing the $W_3$ invariant.

\section*{Acknowledgments}

This work was supported in part
by Deutsche Forschungsgemeinschaft through SFB 652 (project no. B5). 


\end{document}